\documentclass{article}

\usepackage{arxiv}

\usepackage[utf8]{inputenc} 
\usepackage[T1]{fontenc}    
\usepackage{hyperref}       
\usepackage{url}            
\usepackage{booktabs}       
\usepackage{amsfonts}       
\usepackage{nicefrac}       
\usepackage{microtype}      
\usepackage{lipsum}		
\usepackage{graphicx}
\usepackage{natbib}
\usepackage{doi}
\usepackage{graphics}
\usepackage{graphicx}
\usepackage{epsfig}
\usepackage{amssymb}
\usepackage{amsthm}
\usepackage{bm}
\usepackage{amsmath}
\usepackage{color}
\usepackage{array}
\usepackage{todonotes}

\title{Nonlinear Indentation of \\ Second-order Hyperelastic Materials}

\author{ \href{https://orcid.org/0000-0001-6868-4646}{\includegraphics[scale=0.06]{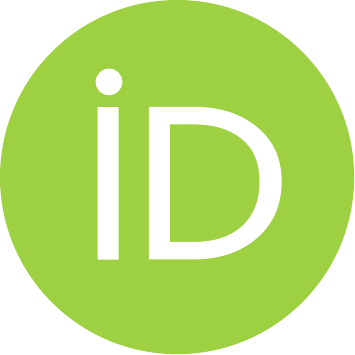}\hspace{1mm}Yangkun Du}\thanks{corresponding author: \texttt{yangkun.du@glasgow.ac.uk} } \\
	School of Mathematics and Statistics\\
	University of Glasgow\\
	Glasgow G12 8QQ, UK\\
	\And \hspace{1mm}Peter Stewart\\
	School of Mathematics and Statistics\\
	University of Glasgow\\
	Glasgow G12 8QQ, UK\\
	\And \hspace{1mm}Nicholas A Hill\\
	School of Mathematics and Statistics\\
	University of Glasgow\\
	Glasgow G12 8QQ, UK\\
	\And \hspace{1mm}Huabing Yin\\
	Biomedical Engineering, School of Engineering\\
	University of Glasgow\\
	Glasgow G12 8LT, UK\\
	\And \hspace{1mm}Raimondo Penta\\
	School of Mathematics and Statistics\\
	University of Glasgow\\
	Glasgow G12 8QQ, UK\\
	\And \hspace{1mm}Jakub K\"{o}ry\\
	School of Mathematics and Statistics\\
	University of Glasgow\\
	Glasgow G12 8QQ, UK\\
	\And \hspace{1mm}Xiaoyu Luo\\
	School of Mathematics and Statistics\\
	University of Glasgow\\
	Glasgow G12 8QQ, UK\\
	\And \hspace{1mm}Raymond Ogden\\
	School of Mathematics and Statistics\\
	University of Glasgow\\
	Glasgow G12 8QQ, UK\\
}

\renewcommand{\shorttitle}{Nonlinear Indentation of Second-order Hyperelastic Materials}

\begin{document}
\maketitle

\begin{abstract}
The classical problem of indentation on an elastic substrate has found new applications in the field of the Atomic Force Microscopy.
However, linearly elastic indentation models are not sufficiently accurate to predict the force-displacement relationship at large indentation depths. 
For hyperelastic materials, such as soft polymers and biomaterials, a nonlinear indentation model is needed.
In this paper, we use second-order elasticity theory to capture larger amplitude deformations and material nonlinearity. 
We provide a general solution for the contact problem for deformations that are second-order in indentation amplitude with arbitrary indenter profiles. 
Moreover, we derive analytical solutions by using either parabolic or quartic surfaces to mimic a spherical indenter.
The analytical prediction for a quartic surface agrees well with finite element simulations using a spherical indenter for indentation depths on the order of the indenter radius.
In particular, the relative error between the two approaches is less than 1\% for an indentation depth equal to the indenter radius, an order of magnitude less than that observed with models which are either first-order in indentation amplitude or those which are second-order in indentation amplitude but with a parabolic indenter profile. 
\end{abstract}


\sloppy
\section{Introduction}
\begin{figure}[h]
\begin{center}
\includegraphics[width=.95\textwidth]{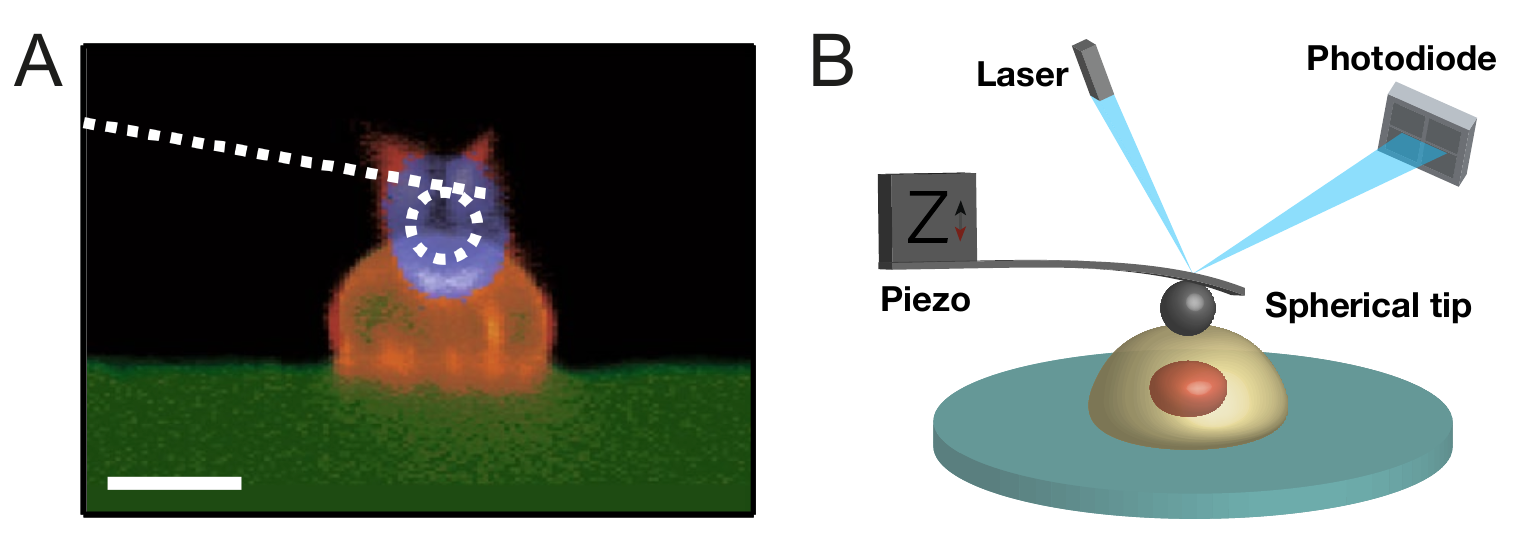}
\caption{(A)Photograph of the nano-indentation test of a microglial cell on polyacrylamide (PAA) substrates by confocal laser scanning microscopy \citep{rheinlaender2020cortical}. (B)The main elements of an AFM setup. The spherical tip interacts with the sample and causes the deflection of the microcantilever. The deflection is then recorded via a laser beam and a four-quadrant photodiode.}
\label{photo-indentation}
\end{center}
\end{figure}

Understanding and quantifying the mechanical characteristics of soft materials, including elastomers and bio-tissues, is of great importance in many engineering applications \citep{chaudhuri2020effects,gensbittel2020mechanical,tian2020mechanical,du2020electro}.
Unlike hard materials, these soft materials, such as hydrogel and cells, are either too fragile or too small to implement traditional macroscopic stretch and compression tests. 
However, the emergence of Atomic Force Microscopy (AFM) (Figure \ref{photo-indentation}) enables us to characterize the mechanical response of these soft materials through local nano-indentation tests \citep{krieg2019atomic,liang2020recent}, measuring the force required to produce a given displacement.
Material constants, such as Young's modulus and the relaxation modulus, can be extracted by fitting the experimental data to theoretical indentation models \citep{chim2018one,efremov2017measuring}.

The classical Hertz model \citep{hertz1881contact} is one of the most widely used theoretical models for elucidating the load-displacement relationship in a frictionless indentation test \citep{johnson1982one}.
Assuming that the contact surface is a small elliptical region while the indentation depth is infinitesimal compared to the scale of the sample, Hertz solved the contact problem by applying the Boussinesq approximation with spatially distributed normal stress \citep{lai2009introduction}.
Over the years, a number of refinements to the Hertz model have been proposed for specific considerations, including substrate effects, friction, adhesive stress, viscoelasticity 
\citep{rheinlaender2020cortical,borodich2014hertz,spence1975hertz,storaakers2005hertz,jin2013adhesive,chim2018one,wang2020characterizing}, and unknown contact conditions for nonlinear materials \citep{chang2018modified}.

In a different approach, \cite{sneddon1965relation} put forward a general analytical solution to the indentation problem in terms of dual-integral functions with an arbitrary indenter profile. 
This solution is consistent with the Hertz model when the indenter is of paraboloid shape.
In addition, Sneddon also provided an analytical expression for the load-displacement relationship in terms of the material constants and contact radius when the indenter has a hemispherical shape.
However, both the Hertz and Sneddon approaches were focused on finding the solution to the Boussinesq problem when the deformations are infinitesimal and the substrate is a linearly elastic half-space.
However, for soft materials, the displacements can become large under moderate loads, and the stress-strain relationship is unlikely to be linearly elastic.
In addition, \cite{zhang2017effects} used finite element (FE) simulations to investigate the impact of large deformations and material nonlinearity on the indentation model of hyperelastic samples. 

Based on their FE simulations, \cite{zhang2014spherical} proposed explicit empirical load-displacement relationships for several hyperelastic materials through a dimensional analysis approach.
Moreover, robust nonlinear indentation models have been applied to materials which exhibit a layered structure \citep{chen2012nanoindentation}, poroelasticity \citep{duan2012effect}, and plasticity \citep{song2013elastic}  by fitting to FE simulations.
However, these numerical-simulation-based finite indentation models are not universally applicable since they rely significantly on the particulars of the FE models, including the material, geometry, mesh, and boundary conditions. 
Therefore, a general theoretical nonlinear indentation model for hyperelastic materials is needed.

From the perspective of mathematical modelling, nonlinear indentation problems are significantly more localized and complicated. 
It seems unlikely that a fully nonlinear finite deformation model can be established for a general nonlinear (or even hyperelastic) material. 
However, we can include larger deformations by extending the linear elasticity approach to higher-order deformation amplitudes as in the weakly nonlinear procedure proposed by \cite{rivlin1953solution}.
In particular, by assuming that the second-order terms in perturbation amplitude induced by products of first-order displacements can introduce additional body forces and surface tractions that can be satisfied by the second-order displacements, Rivlin showed that the second-order nonlinear boundary value problem could be reduced to two linear boundary value problems in classical elasticity theory.

By combining this method with the first-order analytical solution of \cite{sneddon1965relation}, which indeed cannot always be used to describe nanoindentation of non-linear elastic materials \citep{zhang2014applicability}, \cite{sabin1983contact} presented a general solution of the indentation problem up to second-order deformations, using a paraboloid to approximate to the analytical solution for a hemispherical indenter.
Moreover, \cite{giannakopoulos2007spherical} repeated the calculation of \cite{sabin1983contact}, and obtained a different load-displacement relationship by specifying the third-order material constants in terms of the Lam\'{e} constants. 
However, both their analytical approaches exhibited an overestimation of the indentation force compared to  FE simulations and experimental data \citep{liu2010nonlinear}.
 
In this paper, we revisit the second-order indentation problem of \cite{sabin1983contact}, correcting several of their expressions and finding a result that agrees significantly better with the numerical calculations. 
In particular, the second-order nonlinear boundary value problem is reduced to two linear elastic boundary value problems. 
Based on the first-order solution constructed by \cite{sneddon1965relation}, and introducing the integral transform method, the general solutions are expressed in the form of Hankel transforms of potential functions. 
To mimic the spherical indentation more accurately, we further provide asymptotic analytical solutions using a higher-order quartic surface to approximate the spherical indenter. 
We also implement FE simulations to verify our second-order indentation models for incompressible neo-Hookean and Mooney-Rivlin materials.
Finally, we discuss the limitations of this current second-order elasticity method in accounting for more sophisticated incompressible hyperelastic materials.

The paper is organised as follows. First, in Section 2, we recap the method for expanding the governing equation to second-order in indentation depth. Next,  in Section 3, we present the general mathematical modelling of the finite indentation problem. Furthermore, in Section 4, we provide the (corrected) solution up to second-order in indentation amplitude for both parabolic and quartic indenter profiles. Finally, we implement FE simulations to verify the second-order analytical results in Section 5, and make concluding remarks in Section 6.

\section{Second-order elasticity method}
Suppose that an isotropic elastic body undergoes a nonlinear deformation, so that the point $x_i$ is moved to $x_i+u_i$, where $u_i$ is the displacement vector.
We define the deformation gradient tensor 
\begin{equation}
g_{i k}=\left(\delta_{i s}+\frac{\partial u_{i}}{\partial x_{s}}\right)\left(\delta_{k s}+\frac{\partial u_{k}}{\partial x_{s}}\right),
\end{equation}
and its corresponding scalar invariants
\begin{equation}
\mathcal I_{1}=g_{s s}, \quad \mathcal I_{2}=G_{s s}, \quad \mathcal I_{3}=\operatorname{det} g_{i k}.
\end{equation}
Then the stress components can be obtained as
\begin{equation}
t_{i k}=\frac{2}{\tau}\left[g_{i k} \frac{\partial W}{\partial \mathcal I_{1}}-G_{i k} \frac{\partial W}{\partial \mathcal I_{2}}+\left(\mathcal I_{3} \frac{\partial W}{\partial \mathcal I_{3}}+\mathcal I_{2} \frac{\partial W}{\partial \mathcal I_{2}}\right) \delta_{i k}\right],
\label{StressComponents}
\end{equation}
where $\tau=\det(\delta_{i k}+\partial u_i/\partial x_k)$, $W$ is the strain-energy function, $G_{ik }$ is the co-factor matrix of $g_{ik}$, and $\delta_{i k}$ is the Kronecker delta.
We shall assume that the displacement gradients are asymptotically small i.e. $|\partial{u_i}/\partial{x_k}|\sim\varepsilon$, say, where $0<\varepsilon\ll1$, and that the strain energy function is given by the third-order \cite{murnaghan1937finite} expansion
\begin{equation}
W=a_{0} J_{1} +a_{1} J_{2} +a_{2} J_{1}^{2} +a_{3} J_{1} J_{2} +a_{4} J_{1}^{3} +a_{5} J_{3},
\label{MurnaghanEF}
\end{equation}
where $a_0, ..., a_5$ are material constants, and $J_{1}=\mathcal I_1-3$, $J_{2}=\mathcal I_2-2\mathcal I_1+3$, $J_{3}=\mathcal I_3-\mathcal I_2+\mathcal I_1-1$ are three other independent scalar invariants that are respectively $O(\varepsilon), O(\varepsilon^2), \text{and}~ O(\varepsilon^3)$.
In addition, $a_0=0$ if the undeformed configuration is stress-free, while $a_1$ and $a_2$ are related to the Lam\'e constants $\lambda$ and $\mu$ by
\begin{equation}
a_1=-\mu/2, \quad a_2 = (\lambda + 2 \mu)/8,
\end{equation}
and to the Young modulus $E_Y$ and Poisson's ratio $\eta$ by
\begin{equation}
a_{1}=-\frac{1}{4} \frac{E_Y}{1+\eta},\quad a_{2}=\frac{1}{8} E_Y \frac{1-\eta}{(1+\eta)(1-2 \eta)}.
\label{atoYP}
\end{equation}
In \eqref{atoYP}, we suppose that $-1< \eta <1/2$ and  will address the limit  $\eta \rightarrow 1/2$ for an incompressible material below.
From equations \eqref{StressComponents} and \eqref{MurnaghanEF}, the stress components up to second order of quantities $\partial{u_i}/\partial{x_k}$, i.e. $O(\varepsilon^2)$, are 
\begin{equation}
\begin{aligned}
t_{i k}&=2\left[\left\{-a_{1}  e_{i k}+2\left(a_{1}+2 a_{2}\right)  \Delta \delta_{i k}\right\}\right. \\
&~~+\left\{\left(4 a_{2}-2 a_{3}+a_{1}\right) \Delta e_{i k}-a_{1} \alpha_{i k}-\left(a_{1}-a_{5}\right) E_{i k} \right.\\
&~~\left.\left.
+\left(\left(a_{1}+2 a_{2}\right) \alpha+\left(a_{1}+a_{3}\right) E+2\left(6 a_{4}+2 a_{3}-a_{1}-2 a_{2}\right) \Delta^{2}\right) \delta_{i k}\right\}\right],
\end{aligned}
\end{equation}
where 
\begin{align}
e_{i k}=\frac{\partial u_{i}}{\partial x_{k}}+\frac{\partial u_{k}}{\partial x_{i}},~\Delta=e_{s s}/2,~ 
\alpha_{i k}=\frac{\partial u_{i}}{\partial x_{s}} \frac{\partial u_{k}}{\partial x_{s}},~ \alpha=\alpha_{ss},~ E=E_{ss},
\end{align}
and $E_{ik}$ is the cofactor matrix of $e_{ik}$.

Furthermore, up to $O(\varepsilon^2)$, we can expand the displacement field as 
\begin{equation}
u_{i}= v_{i}+w_{i},
\label{ui}
\end{equation}
where the $v_{i}=O(\varepsilon)$ and $w_{i}=O(\varepsilon^2)$.
Hence, the stress to $O(\varepsilon^2)$ can be separated as
\begin{equation}
t_{ik}=\tau_{ik}+\tau_{i k}^\prime+\tau_{i k}^{\prime\prime},
\label{tik}
\end{equation}
where
\begin{equation}
\begin{aligned}
\tau_{i k}&=2\left[-a_{1} e_{i k}^\prime+2\left(a_{1}+2 a_{2}\right) \Delta^\prime \delta_{i k}\right]\\
\end{aligned}
\label{StressS1}
\end{equation}
is the first-order stress component,
\begin{equation}
\begin{aligned}
\tau_{i k}^\prime=&2\left[\left(4 a_{2}-2 a_{3}+a_{1}\right) \Delta^\prime e_{i k}^\prime-a_{1} \alpha_{i k}^\prime-\left(a_{1}-a_{5}\right) E_{i k}^\prime\right. \\
&\left.~+\left\{\left(a_{1}+2 a_{2}\right) \alpha^\prime+\left(a_{1}+a_{3}\right) E^\prime+2\left(6 a_{4}+2 a_{3}-a_{1}-2 a_{2}\right) \Delta^{\prime 2}\right\} \delta_{i k}\right], \\
\tau_{i k}^{\prime\prime}=&2\left[-a_{1} e_{i k}^{\prime\prime}+2\left(a_{1}+2 a_{2}\right)\Delta^{\prime\prime} \delta_{i k}\right],
\end{aligned}
\label{StressS2}
\end{equation}
are the second-order stress components,
and 
\begin{equation}
\begin{aligned}
&e_{i k}^\prime=\frac{\partial v_{i}}{\partial x_{k}}+\frac{\partial v_{k}}{\partial x_{i}},~\Delta^\prime=e_{s s}^\prime/2,~ e_{i k}^{\prime\prime}=\frac{\partial w_{i}}{\partial x_{k}}+\frac{\partial w_{k}}{\partial x_{i}},~\Delta^{\prime\prime}=e_{s s}^{\prime\prime}/2,\\
&\alpha_{i k}^\prime=\frac{\partial v_{i}}{\partial x_{s}} \frac{\partial v_{k}}{\partial x_{s}},~ \alpha^\prime=\alpha_{ss}^\prime,~E=E_{ss}^\prime,~\text{and}~  E_{ik}^\prime \text{ is the cofactor matrix of } e_{ik}^\prime.
\end{aligned}
\label{StrainS}
\end{equation}
Then, the equilibrium equation and the boundary conditions to $O(\varepsilon^2)$ are 
\begin{equation}
\begin{aligned}
&\frac{\partial \tau_{i k}}{\partial x_{k}}+   \frac{\partial \tau_{i k}^{\prime \prime}}{\partial x_{k}}+\left[\Delta^{\prime} \delta_{s k}-\frac{\partial v_{s}}{\partial x_{k}}\right] \frac{\partial \tau_{i k}}{\partial x_{s}}+\frac{\partial \tau_{i k}^{\prime}}{\partial x_{k}}+\varrho_{0}  X_{i}=0, \\
&\mathcal T_{i}=\varepsilon l_{k}\tau_{i k}+ \left[\Delta^{\prime} \delta_{s k}-\frac{\partial v_{s}}{\partial x_{k}}\right] l_{s} \tau_{i k}+l_{k}\left(\tau_{i k}^{\prime}+\tau_{i k}^{\prime \prime}\right),
\end{aligned}
\end{equation}
where $X_i$  and $\mathcal T_i$ are the body force and surface traction associated with the first-order displacement $v_i$, respectively.
Following \cite{rivlin1953solution}, the second-order terms of the equilibrium equation and the boundary condition induced by the first-order displacement $v_i$ can be considered as an additional body force $X_{i}^{\prime } $  and the surface traction $\mathcal T_i$, respectively,
\begin{equation}
\begin{aligned}
& \rho _{0} X_{i}^{\prime } =\left[ \Delta ^{\prime } \delta _{sk} -\frac{\partial v_{s}}{\partial x_{k}}\right]\frac{\partial \tau _{ik}}{\partial x_{s}} +\frac{\partial \tau _{ik}^{\prime }}{\partial x_{k}},\\
& \mathcal  T_{i}^{\prime}= -\left[\Delta \delta_{s k}-\frac{\partial v_{s}}{\partial x_{k}}\right] l_{s} \tau_{i k} - l_{k} \tau_{i k}^{\prime},
\end{aligned}
\end{equation}
where $l_k$ are the direction-cosines of the normal to the deformed surface of the body.
The additional body force and surface traction give rise to a second-order deformation.
Thus, the equilibrium equation and the boundary condition of this second-order elastic problem can be reduced to two linear elastic problems of at $O(\varepsilon)$ and $O(\varepsilon^2)$, respectively,
\begin{equation}
\begin{aligned}
&\frac{\partial \tau _{ik}}{\partial x_{k}} +\rho _{0} X_{i} =0\text{~and~}\mathcal T_{i}=l_k\tau_{ik},\\
&\frac{\partial \tau _{ik}^{''}}{\partial x_{k}} +\rho _{0}X_{i}^{\prime } =0 \text{~and~} \mathcal T_{i}^{\prime}=l_k\tau_{ik}^{''}.
\label{EQbm}
\end{aligned}
\end{equation}

\section{Mathematical modelling of the nonlinear indentation}
\begin{figure}[h]
\begin{center}
\includegraphics[width=0.6\textwidth]{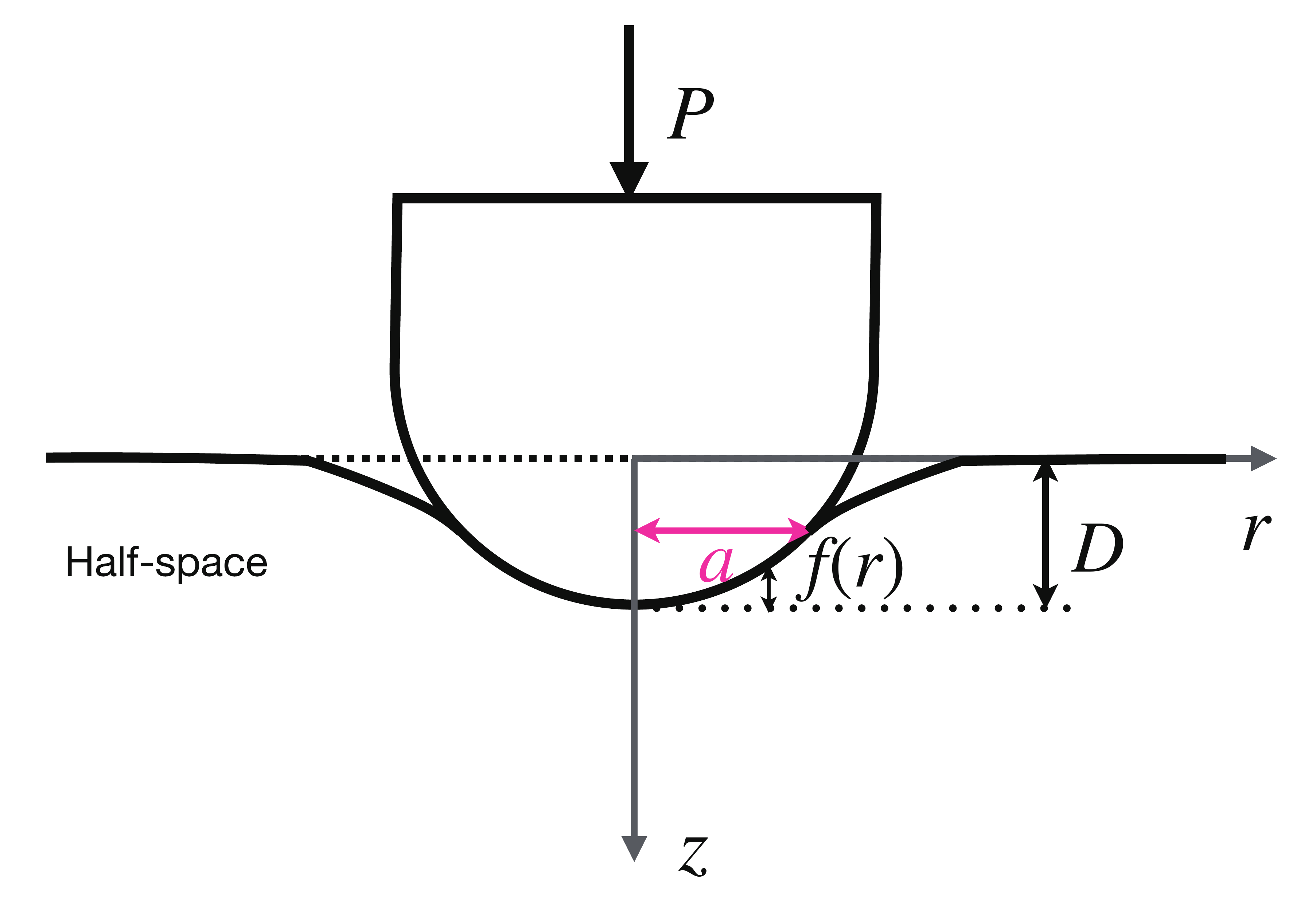}
\caption{The deformed configuration of the nonlinear indentation by an axisymmetric curved indenter.}
\label{fig1}
\end{center}
\end{figure}

As shown in Figure \ref{fig1}, suppose that a hyperelastic half-space body is approached by a rigid axisymmetric indenter with an arbitrary profile $f(r)$. The deformed half-space is defined in terms of the cylindrical coordinates $(r,\theta,z)$ centred with the indenter. The contact radius $a$ is fixed, and we shall determine the corresponding at indentation depth $D$, see Figure \ref{fig1}. 
In addition, we assume that there is no internal body force within the half-space, and that the interface is frictionless. 
Hence, at the contact surface $z=0$ in the deformed configuration, we have the boundary conditions
\begin{equation}
\begin{aligned}
& u_z(r,0)=D-f(r) \quad(0\leq r\leq a),\\
& t_{zz}(r,0)=0\quad (r > a),\quad t_{rz}(r,0)=0  \quad(r\geq 0).
\label{BCs}
\end{aligned}
\end{equation}
We further assume that radius of curvature of the tip of the axisymmetric indenter is $R\gg a$, and thus set $\varepsilon= a/R$ as the small parameter.
Similar to the expansion of the displacement and the stress fields, the indentation depth can be also expanded as 
\begin{equation}
D= D_1 + D_2,
\end{equation}
where $D_1=O(\varepsilon)$ and $D_2=O(\varepsilon^2)$, and $0\leq f(r) \leq D$.
Hence, the boundary conditions \eqref{BCs} of the first-order $O(\varepsilon)$ deformation are
\begin{equation}
\begin{aligned}
& v_z(r,0)=D_1- f(r)  \quad (0\leq r\leq a),\\
&  \tau_{zz}(r,0)=0\quad (r > a),\quad   \tau_{rz}(r,0)=0 \quad (r\geq 0),
\label{BCs1}
\end{aligned}
\end{equation}
and those of the second-order $O(\varepsilon^2)$ deformation  are 
\begin{equation}
\begin{aligned}
&  w_z(r,0)= D_2 \quad (0\leq r\leq a),\\
& \tau_{zz}^{\prime\prime}(r,0)=- \mathcal T_{z}^\prime \quad (r > a),\quad  \tau_{rz}^{\prime\prime}(r,0)=- \mathcal T_{r}^\prime \quad (r\geq 0).
\label{BCs2}
\end{aligned}
\end{equation}
Here we assume that the shape of the indenter (e.g. spherical) is such that $f(r)$ does not contribute terms at $O(\varepsilon^2)$.
Following \cite{rivlin1953solution}, we solve this second-order contact problem in two steps.
First, we obtain the first-order solution that satisfies the boundary condition \eqref{BCs1} based on classical elasticity theory. Second, having calculated additional second-order body force and surface traction from the first-order solution, we find the solution satisfies the boundary condition \eqref{BCs2}, also making use of classical elasticity theory.

In the following subsections, we present the complete derivation process for the general solution of the second-order indentation problem, referencing and correcting \cite{sabin1983contact}'s work. 
For better understanding and consistency, we adopt consistent notation and provide more details.

\subsection{The first-order solution}
\cite{sneddon1965relation} constructed a general analytical solution of the first-order contact problem with the equilibrium equation \eqref{EQbm}$_1$ and boundary condition \eqref{BCs1}, using an integral transform method. 
The solution of the first-order displacement is
\begin{equation}
\begin{aligned}
&v_{r} (r,z)=-(1-2\eta )G_{0} (r,z) +zG_{1} (r,z) ,\\
&v_{z} (r,z)=2(1-\eta )F_{0} (r,z) +zF_{1} (r,z),
\end{aligned}
\end{equation}
where 
\begin{equation}
\begin{aligned}
&F_{i}(r, z)=H_{0}\left[\xi^{-1+i} \psi(\xi) \mathrm{e}^{-\xi z} ; \xi \rightarrow r\right],
~G_{i}(r, z)=H_{1}\left[\xi^{-1+i} \psi(\xi) \mathrm{e}^{-\xi z} ; \xi \rightarrow r\right],\\
&\psi(\xi)=\frac{1}{2(1-\eta)} \int_{0}^{a} \chi(t) \cos (\xi t) \mathrm{d} t, 
~\chi(t)=\frac{2 D_{1}}{\pi}-\frac{2 t}{\pi} \int_{0}^{t} \frac{f^{\prime}(x)}{\sqrt{t^{2}-x^{2}}} \mathrm{~d} x,
\label{FGpsichi}
\end{aligned}
\end{equation}
and $H_{j}[g(x) ; x \rightarrow \zeta]=\int_{0}^{\infty} x g(x) J_{j}(\zeta x) \mathrm{d} x$ is the $j^{th}$-order Hankel transform of function $g(x)$.
In addition, following \cite{sneddon1965relation}, $\chi(a)=0$ is required to make sure that  $\tau_{zz}(a,0)$ tends to a finite limit.
Then, combining the equations \eqref{atoYP}, \eqref{StressS1}, \eqref{StrainS}, and \eqref{FGpsichi}, we obtain the first-order stress components
\begin{equation}
\begin{aligned}
&\tau _{rr} (r,z)=-2\mu \left[ F_{1}(r,z) -zF_{2}(r,z) -(1-2\eta )\frac{G_{0}(r,z)}{r} +\frac{zG_{1}(r,z)}{r}\right],\\
&\tau _{\theta \theta } (r,z)=-2\mu \left[ 2\eta F_{1}(r,z) +(1-2\eta )\frac{G_{0}(r,z)}{r} -\frac{zG_{1}(r,z)}{r}\right],\\
&\tau _{zz} (r,z)=-2\mu \left[F_{1}(r,z) +zF_{2}(r,z)\right] ,\ \ \tau _{rz} (r,z)=-2\mu zG_{2}(r,z).
\end{aligned}
\end{equation}
Moreover, according to equation \eqref{StressS2}$_2$, the second-order stress components at the contact surface $z=0$ induced by the first-order deformation are given by
\begin{equation}
\begin{aligned}
\tau_{rr}^\prime(r,0)=&2 (1-2 \eta )^2 \left[a_{1}+4 (a_{2}+3 a_{3}+12 a_{4})\right] F_{1}^2(r,0) \\
&-8 (1-2 \eta )^2 (a_1+5 a_2-3 a_3-a_5) \frac{F_{1}(r,0) G_{0}(r,0)}{r}\\
&-2 (1-2 \eta )^2 (3 a_1-4 a_2+4 a_3)\frac{G_0^2(r,0)}{r^2}+8 (1-\eta)^2 (a_{1}+4 a_{2}) G_{1}^2(r,0),\\
\tau_{\theta\theta}^\prime(r,0)=&\frac{2 (1-2 \eta )^2}{r^2} \left[2 (-3 a_1+2 (5 a_3+12 a_4+a_5-3 a_2)) r^2 F_1^2(r,0)\right.\\
&+2 (5 a_1+6 a_2-2 (a_3+a_5)) r F_1(r,0) G_0(r,0)\\
&\left.+(4 a_2-4 a_3-3 a_1) G_0^2(r,0)\right],\\
\tau_{zz}^\prime(r,0)=&2 (1-2 \eta )^2 \left[a_{1}+4 (a_{2}+3 a_{3}+12 a_{4})\right] F_{1}^2(r,0) \\
&+4 (1-2 \eta )^2 (a_{1}+2 a_{2}-2 a_{3}-2 a_{5})\frac{G_{0}^2(r,0)-r F_{1}(r,0) G_{0}(r,0)}{r^2}\\
&+8 (1-\eta)^2 (a_{1}+4 a_{2}) G_{1}^2(r,0),\\
\tau _{rz}^{\prime } (r,0)= & \frac{4 a_1 }{r} (1-2\eta )(1-\eta )G_{1} (r,0)G_{0} (r,0).
\end{aligned}
\label{AddBF}
\end{equation}
The additional body force, $\varrho_0 X_r^\prime$, and the surface traction, $\mathcal T_{i}^\prime$, are given by
\begin{equation}
\begin{aligned}
\varrho _{0} X_{r}^{\prime } = & \frac{2}{r} (1-2\eta )^{2}( a_{1} -4a_{2} +4a_{3})\left[ F_{1} (r,0)-\frac{2}{r} G_{0} (r,0)\right]^{2}\\
 & +2(1-2\eta )G_{2} (r,0)\{[ (1-2\eta )( 3a_{1} +20a_{2} -12a_{3} -4a_{5})\\
 & +2( 3a_{1} -2a_{5})]\frac{G_{0} (r,0)}{r} +2[( a_{1} +8a_{2} -4a_{3})\\
 & -(1-2\eta )( a_{1} +4a_{2} +12a_{3} +48a_{4})] F_{1} (r,0)\}\\
 & -8(1-\eta )G_{1} (r,0)\left[( a_{1} -4a_{2}) F_{2} (r,0)+\frac{4a_{2}}{r} G_{1} (r,0)\right],\\
\mathcal T_{r}^{\prime } = & -4\mu (1-\eta )G_{1}(r,0)\left[ F_{1} (r,0)-(1-2\eta )\frac{G_{0}(r,0)}{r}\right] +\tau _{rz}^{\prime } (r,0),\\
\mathcal T_{\theta }^{\prime } = & 0,\quad \mathcal T_{z}^{\prime } = 2\mu (1-2\eta )F_{1}^{2} (r,0)+\tau _{zz}^{\prime } (r,0).
\end{aligned}
\label{AddST}
\end{equation}

\subsection{The second-order solutions}
We have calculated the additional body force and the surface traction \eqref{AddST} based on the first-order solutions. Now, we construct the solution of the second linear problem with the equilibrium equation \eqref{EQbm}$_2$ and boundary condition \eqref{BCs2}.

The equilibrium equation \eqref{EQbm}$_2$ can be expanded to give
\begin{equation}
\begin{aligned}
&\frac{\partial \tau_{r r}^{\prime \prime}}{\partial r}+\frac{\partial \tau_{r z}^{\prime \prime}}{\partial z}+\frac{\tau_{r r}^{\prime \prime}-\tau_{\theta \theta}^{\prime \prime}}{r}+\varrho_{0} X_{r}^{\prime}=0,\\
&\frac{\partial \tau_{r z}^{\prime \prime}}{\partial r}+\frac{\partial \tau_{z z}^{\prime \prime}}{\partial z}+\frac{\tau_{r z}^{\prime \prime}}{r} +\varrho_{0} X_{z}^{\prime}=0.
\end{aligned}
\label{EQsecond}
\end{equation}
Since the system \eqref{EQsecond} is linear, the solution can be decomposed into the sum of three separate linear problems in terms of the stresses $\sigma_{ik}^{\prime}$, $\sigma_{ik}^{\prime\prime}$, $\sigma_{ik}^{\prime\prime \prime}$ and the displacements $w_{i}^{\prime}$, $w_{i}^{\prime\prime}$, $w_{i}^{\prime\prime\prime}$, written as
\begin{equation}
\tau_{ik}^{\prime \prime}=\sigma_{ik}^{\prime}+\sigma_{ik}^{\prime\prime}+\sigma_{ik}^{\prime\prime\prime},\quad
w_{i}=w_{i}^{\prime}+w_{i}^{\prime\prime}+w_{i}^{\prime\prime\prime}.
\label{tauik&wi}
\end{equation}
Thus, the equilibrium equation \eqref{EQsecond} and the boundary condition \eqref{BCs2} are equivalent to the sum of linear problem $(i)$,
\begin{equation}
\left\{\begin{aligned}
&\frac{\partial \sigma_{r r}^{\prime}}{\partial r}+\frac{\partial \sigma_{r z}^{\prime }}{\partial z}+\frac{\sigma_{r r}^{ \prime}-\sigma_{\theta \theta}^{\prime }}{r}+\varrho_{0} X_{r}^{\prime}=0\\
&\frac{\partial \sigma_{r z}^{\prime }}{\partial r}+\frac{\partial \sigma_{z z}^{\prime }}{\partial z}+\frac{\sigma_{r z}^{\prime }}{r} +\varrho_{0} X_{z}^{\prime}=0
\end{aligned}\right.~
\left\{\begin{aligned}
& w_z^{\prime}(r,0)=0 \quad (0\leq r\leq a)\\
& \sigma_{zz}^{\prime}(r,0)=-\mathcal T_{z}^\prime\quad (r > a)\\
& \sigma_{rz}^{\prime}(r,0)=0 \quad (r\geq 0)
\end{aligned}\right.,
\label{EQsecond-1}
\end{equation}
linear problem $(ii)$,
\begin{equation}
\left\{\begin{aligned}
&\frac{\partial \sigma_{r r}^{\prime\prime}}{\partial r}+\frac{\partial \sigma_{r z}^{\prime \prime}}{\partial z}+\frac{\sigma_{r r}^{ \prime\prime}-\sigma_{\theta \theta}^{\prime\prime }}{r}=0\\
&\frac{\partial \sigma_{r z}^{\prime\prime }}{\partial r}+\frac{\partial \sigma_{z z}^{\prime\prime }}{\partial z}+\frac{\sigma_{r z}^{\prime\prime}}{r} =0
\end{aligned}\right.~
\left\{\begin{aligned}
& w_z^{\prime\prime}(r,0)=0 \quad (0\leq r\leq a)\\
& \sigma_{zz}^{\prime\prime}(r,0)=0 \quad (r > a)\\
& \sigma_{rz}^{\prime\prime}(r,0)=-\mathcal T_{r}^\prime \quad (r\geq 0)
\end{aligned}\right.,
\label{EQsecond-2}
\end{equation}
and linear problem $(iii)$,
\begin{equation}
\left\{\begin{aligned}
&\frac{\partial \sigma_{r r}^{\prime\prime\prime}}{\partial r}+\frac{\partial \sigma_{r z}^{\prime\prime\prime }}{\partial z}+\frac{\sigma_{r r}^{ \prime\prime\prime}-\sigma_{\theta \theta}^{\prime\prime\prime }}{r}=0\\
&\frac{\partial \sigma_{r z}^{\prime\prime\prime }}{\partial r}+\frac{\partial \sigma_{z z}^{\prime\prime\prime }}{\partial z}+\frac{\sigma_{r z}^{\prime\prime\prime}}{r} =0
\end{aligned}\right.~
\left\{\begin{aligned}
& w_z^{\prime\prime\prime}(r,0)=D_2 \quad (0\leq r\leq a)\\
& \sigma_{zz}^{\prime\prime\prime}(r,0)=0 \quad (r > a)\\
& \sigma_{rz}^{\prime\prime\prime}(r,0)=0 \quad (r\geq 0)
\end{aligned}\right..
\label{EQsecond-3}
\end{equation}
This separation helps to simplify the calculations. 
Furthermore, these linear elastic problems can be solved by using Papkovitch--Neuber potential function method \citep{lai2009introduction}. The general solution of the displacement vector for linear elastostatic problems is given by 
\begin{equation}
\boldsymbol{w}=\nabla(\varphi+\boldsymbol{x} \cdot \boldsymbol \psi)-4(1-\eta) \boldsymbol{\psi},
\label{solution-2}
\end{equation}
where $\varphi$ is a scalar function, and $\boldsymbol{\psi}$ is a vector function. 
With the decomposition \eqref{solution-2}, the general equilibrium equation \eqref{EQbm}$_2$ can be rewritten as 
\begin{equation}
\frac{2 \mu(1-\eta)}{(1-2 \eta)}\left(x_{n} \frac{\partial \nabla^{2} \psi_{n}}{\partial x_{i}}+\nabla^{2} \psi_{i}+\frac{\partial \nabla^{2} \varphi}{\partial x_{i}}\right)-4 \mu(1-\eta) \nabla^{2} \psi_{i}+\varrho^\prime X_{i}=0.
\label{EQinPF}
\end{equation}
For this axisymmetric problem in cylindrical coordinates, the two potential functions can be specified as $\varphi=\varphi(r,z)$ and $\boldsymbol{\psi}=(0,0,\psi(r,z))$. Hence, the equilibrium equation \eqref{EQbm}$_2$ and \eqref{EQinPF} are equivalent to 
\begin{equation}
\begin{aligned}
&\nabla^{2} \psi=\frac{\varrho_{0}}{4 \mu(1-\eta)}\left[X_z^{\prime}+\frac{\partial S}{\partial z} \right] \equiv K_{1}(r,z),\\
&\nabla^{2} \varphi=\frac{\varrho_{0}}{4 \mu(1-\eta)}\left[-z \frac{\partial S}{\partial z}+2(1-2 \eta) S-z X_{z}^{\prime}\right]\equiv K_{2}(r,z),
\end{aligned}
\end{equation}
where $S(r,z)=\int_r^\infty X_r^\prime(r^\prime,z)\mathrm d r^\prime$. Moreover, from \cite{sabin1983contact}, the two potential functions are given by
\begin{equation}
\begin{aligned}
\psi(r,z)=H_{0}\left[\left(B(\xi)+K_1^{ *} (\xi, z)\right) \mathrm{e}^{-\xi z} ; \xi \rightarrow r\right],\\
\varphi(r,z)=H_{0}\left[\left(A(\xi)+K_2^{ *} (\xi, z)\right) \mathrm{e}^{-\xi z} ; \xi \rightarrow r\right],
\label{Hankel-phi-psi}
\end{aligned}
\end{equation}
where $A(\xi)$ and $B(\xi)$ are arbitrary functions that need to be determined from the boundary conditions, and 
\begin{equation}
\begin{aligned}
K_1^{ *}(\xi, z)=\int_{0}^{z} \mathrm{e}^{2 \xi z_{2}} \int_{0}^{z_{2}} \mathrm{e}^{-\xi z_{1}} \int _{0}^{\infty } rK_{1}( r, z_{1}) J_0( \xi r)~ \mathrm{d} r \mathrm{d} z_{1} \mathrm{d} z_{2},\\
K_2^{ *}(\xi, z)=\int_{0}^{z} \mathrm{e}^{2 \xi z_{2}} \int_{0}^{z_{2}} \mathrm{e}^{-\xi z_{1}} \int _{0}^{\infty } rK_{2}( r, z_{1}) J_0( \xi r) ~ \mathrm{d} r \mathrm{d} z_{1} \mathrm{d} z_{2}.
\end{aligned}
\end{equation}
Then, combining Eqs. \eqref{StressS1}, \eqref{StressS2}, \eqref{StrainS}, and \eqref{solution-2}, we obtain the general solutions of the second-order displacement $\boldsymbol{w}$ and the corresponding stress components as
\begin{equation}
\begin{aligned}
&w_{z}=\frac{\partial \varphi}{\partial z}+z \frac{\partial \psi}{\partial z}-(3-4 \eta) \psi,\quad
\tau_{r z}^{\prime \prime}=2 \mu \frac{\partial}{\partial r}\left[\frac{\partial \varphi}{\partial z}+z \frac{\partial \psi}{\partial z}-(1-2 \eta) \psi\right], \\
&\tau_{z z}^{\prime \prime}=2 \mu\left[\frac{\partial^{2} \varphi}{\partial z^{2}}+z \frac{\partial^{2} \psi}{\partial z^{2}}-2(1-\eta) \frac{\partial \psi}{\partial z}+\frac{\eta}{1-2 \eta}\left(\nabla^{2} \varphi+z \nabla^{2} \psi\right)\right].
\end{aligned}
\label{SolutionSecondLinear}
\end{equation}
Note that \eqref{SolutionSecondLinear} is the general solution for the three separate linear problems \eqref{EQsecond-1}, \eqref{EQsecond-2}, and \eqref{EQsecond-3}. Next, we derive the functions $A(\xi)$ and $B(\xi)$ in \eqref{Hankel-phi-psi} by applying the boundary conditions to each of the systems $(i)$, $(ii)$, and $(iii)$.

\subsubsection{Solving linear problem $(i)$}
\label{3.2.1}

First, based on the equation \eqref{Hankel-phi-psi}, we construct the partial derivatives of the potential functions $\varphi(r,z)$ and $\psi(r,z)$ at the contact surface $z=0$ given by
\begin{equation}
\begin{aligned}
\left.\frac{\partial \psi}{\partial z}\right|_{z=0}=&-H_{0}[\xi B(\xi) ; \xi \rightarrow r],\left.\quad \frac{\partial \psi}{\partial r}\right|_{z=0}=-H_{1}[\xi B (\xi); \xi \rightarrow r], \\
\left.\frac{\partial^{2} \psi}{\partial r^{2}}\right|_{z=0}=&-H_{0}\left[\xi^{2} B(\xi); \xi \rightarrow r\right]+\frac{1}{r} H_{1}[\xi B(\xi) ; \xi \rightarrow r],\\
\left.\frac{\partial^{2} \psi}{\partial z^{2}}\right|_{z=0}=&H_{0}\left[\xi^{2} B(\xi), \xi \rightarrow r\right]+H_{0}\left[\int _{0}^{\infty } r^\prime K_1( r^\prime, 0) J( \xi r^\prime)\mathrm{d} r^\prime, \xi \rightarrow r\right],\\
\left.\frac{\partial^{2} \psi}{\partial r \partial z}\right|_{z=0}=&H_{1}\left[\xi^{2} B(\xi) ; \xi \rightarrow r\right],
\end{aligned}
\label{psi_Derivative}
\end{equation}
and
\begin{equation}
\begin{aligned}
\left.\frac{\partial \varphi}{\partial z}\right|_{z=0}=&-H_{0}[\xi A(\xi) ; \xi \rightarrow r],\left.\quad \frac{\partial \varphi}{\partial r}\right|_{z=0}=-H_{1}[\xi A(\xi) ; \xi \rightarrow r], \\
\left.\frac{\partial^{2} \varphi}{\partial r^{2}}\right|_{z=0}=&-H_{0}\left[\xi^{2} A(\xi) ; \xi \rightarrow r\right]+\frac{1}{r} H_{1}[\xi A(\xi) ; \xi \rightarrow r],\\
\left.\frac{\partial^{2} \varphi}{\partial z^{2}}\right|_{z=0}=&H_{0}\left[\xi^{2} A(\xi), \xi \rightarrow r\right]+H_{0}\left[\int _{0}^{\infty } r^\prime K_2( r^\prime, 0) J( \xi r^\prime)\mathrm{d} r^\prime, \xi \rightarrow r\right],\\
\left.\frac{\partial^{2} \varphi}{\partial r \partial z}\right|_{z=0}=&H_{1}\left[\xi^{2} A(\xi) ; \xi \rightarrow r\right].
\end{aligned}
\label{phi_Derivative}
\end{equation}
For the linear problem $(i)$ of \eqref{EQsecond-1}, based on \eqref{SolutionSecondLinear}, \eqref{psi_Derivative}, and \eqref{phi_Derivative}, we have 
\begin{equation}
\begin{aligned}
\left.w_{z}^{\prime}\right|_{z=0}=&-H_{0}[\xi A(\xi)+(3-4 \eta) B(\xi) ; \xi \rightarrow r]=0 \quad (0\leq r\leq a), \\
\left.\sigma_{z z}^{\prime}\right|_{z=0}=&2 \mu H_{0}\left[\xi^{2} A(\xi)+2(1-\eta) \xi B(\xi) ; \xi \rightarrow r\right]+\varrho_{0} \int_{r}^{\infty} X_{r}^{\prime}\left(r^{\prime}, 0\right) \mathrm{d} r\\
=&-\mathcal T_{z}^\prime\quad (r > a),\\
\left.\sigma_{r z}^{\prime}\right|_{z=0}=&2 \mu \xi\left\{H_{1}[\xi A(\xi)+(1-2 \eta) B(\xi) ; \xi \rightarrow r]\right\}=0 \quad (r\geq 0).
\end{aligned}
\label{sigmaPz0_i}
\end{equation}
The Hankel transform of \eqref{sigmaPz0_i}$_3$ yields
\begin{equation}
\begin{aligned}
B(\xi)=-\frac{\xi A(\xi)}{1-2 \eta}.
\end{aligned}
\label{AB_relationship_i}
\end{equation}
Then, the boundary condition of displacement $w_z^{\prime}$ and the normal stress $\sigma_{zz}^{\prime}$ in \eqref{sigmaPz0_i} can be rewritten as 
\begin{equation}
\begin{aligned}
\left.w_{z}^\prime\right|_{z=0}=&\frac{2(1-\eta)}{1-2 \eta} H_{0}[\xi A(\xi) ; \xi \rightarrow r]=0 \quad (0\leq r\leq a),\\
\left.\sigma_{z z}^{ \prime}\right|_{z=0}=&-\frac{2 \mu}{1-2 \eta} H_{0}\left[\xi^{2} A(\xi) ; \xi \rightarrow r\right]+\varrho_{0} \int_{r}^{\infty} X_{r}^{\prime}\left(r^{\prime}, 0\right) \mathrm{d} r \\
=&-\mathcal T_{z}^\prime\quad (r > a),
\end{aligned}
\end{equation}
from which we can further derive the governing equation of $A(\xi)$ in the form
\begin{equation}
\begin{aligned}
H_{0}[\xi A(\xi) ; \xi \rightarrow r]=&0 \quad (0 \leq r \leq a),\\
H_{0}\left[\xi^{2} A (\xi); \xi \rightarrow r\right]=&\frac{1-2 \eta} {2 \mu}\left[\mathcal T_{ z}^{\prime}+\varrho_{0} \int_{r}^{\infty} X_{r}^{\prime}\left(r^{\prime}, 0\right) \mathrm{d} r\right] \quad (r > a).
\end{aligned}
\label{governingA}
\end{equation}
Following \cite{sneddon1960elementary}, Eq. \eqref{governingA} can be satisfied if 
\begin{equation}
\begin{aligned}
&A(\xi)=\xi^{-2} \int_{a}^{\infty} \beta(t) \cos (\xi t)~\mathrm{d} t, \\
&\beta(t)=\frac{2}{\pi} \frac{1-2 \eta}{2 \mu} \int_{t}^{\infty}\left[\frac{r \mathcal T_{ z}^{\prime}(r, 0)}{\sqrt{r^{2}-t^{2}}}+\varrho_{0} X_{r}^{\prime}(r, 0) \sqrt{r^{2}-t^{2}}\right] \mathrm{d} r.
\end{aligned}
\label{solutionA_i}
\end{equation}
Hence, by combining Eqs. \eqref{sigmaPz0_i}, \eqref{AB_relationship_i}, and \eqref{solutionA_i}, the final solutions for $w_z^\prime$ and $\sigma_{z z}^{\prime}$ of the linear problem $(i)$ in Eq. \eqref{EQsecond-1} are
\begin{equation}
\begin{aligned}
\left.w_{z}^{\prime}\right|_{z=0}=&
 \begin{cases}
 0, &\quad (0 \leq r \leq a), \\
 \dfrac{2(1-\eta)}{1-2 \eta} \int_{a}^{r} \dfrac{\beta(t)}{\sqrt{r^{2}-t^{2}}} \mathrm{~d} t, &\quad (r > a),
 \end{cases}\\
 \left.\sigma_{z z}^{\prime}\right|_{z=0}=&
\begin{cases}\dfrac{2 \mu}{1-2 \eta} \int_{a}^{\infty} \dfrac{t \beta(t)}{\left(t^{2}-r^{2}\right)^{3 / 2}} \mathrm{~d} t+\varrho_{0} \int_{r}^{\infty} X_{r}^{\prime}\left(r^{\prime}, 0\right) \mathrm{d} r^\prime, &\quad (0\leq r \leq a), \\
 -\int_{r}^{\infty} \dfrac{\beta^{\prime}(t)}{\sqrt{t^{2}-r^{2}}} \mathrm{~d} t=-\mathcal T_{z}^\prime, &\quad (r>a).
 \end{cases}
\end{aligned}
\end{equation}

\subsubsection{Solving linear problem $(ii)$}
The linear problem $(ii)$ of Eq. \eqref{EQsecond-2} can be solved by the same approach as in Section \ref{3.2.1} for linear problem $(i)$.
First, based on \eqref{SolutionSecondLinear}, \eqref{psi_Derivative}, and \eqref{phi_Derivative}, we have 
\begin{equation}
\begin{aligned}
\left.w_{z}^{\prime}\right|_{z=0}=&-H_{0}[\xi A(\xi)+(3-4 \eta) B(\xi) ; \xi \rightarrow r]=0 \quad (0\leq r\leq a), \\
\left.\sigma_{z z}^{\prime}\right|_{z=0}=&2 \mu H_{0}\left[\xi^{2} A(\xi)+2(1-\eta) \xi B(\xi) ; \xi \rightarrow r\right]=0\quad (r > a),\\
\left.\sigma_{r z}^{\prime}\right|_{z=0}=&2 \mu \xi\left\{H_{1}[\xi A(\xi)+(1-2 \eta) B(\xi) ; \xi \rightarrow r]\right\}=-\mathcal T_{r}^\prime \quad (r\geq 0).
\end{aligned}
\label{sigmaPz0_ii}
\end{equation}
According to Eq. \eqref{sigmaPz0_ii}$_3$, we have
\begin{equation}
\begin{aligned}
B &=-\frac{\xi A}{(1-2 \eta)}-\frac{\xi^{-1}}{2 \mu(1-2 \eta)} Q(\xi),
\end{aligned}
\label{AB_relationship_ii}
\end{equation}
where $Q(\xi)=H_{1}\left[\mathcal T_{ r}^{\prime}, r \rightarrow \xi\right]$.
Then, the boundary condition for the displacement $w_z^{\prime\prime}$ and the normal stress $\sigma_{zz}^{\prime\prime}$ in Eq. \eqref{sigmaPz0_ii} can be rewritten as
\begin{equation}
\begin{aligned}
\left.w_{z}^{\prime\prime}\right|_{z=0}=&\frac{2(1-\eta)}{1-2 \eta} H_{0}[\xi A ; \xi \rightarrow r]-\frac{3-4 \eta}{2 \mu(1-2 \eta)} H_{0}\left[\xi^{-1} Q(\xi) ; \xi \rightarrow r\right]=0 \quad (0\leq r\leq a),\\
\left.\sigma_{z z}^{\prime \prime}\right|_{z=0}=&\frac{-2 \mu}{1-2 \eta} H_{0}\left[\xi^{2} A ; \xi \rightarrow r\right]+\frac{2(1-\eta)}{1-2 \eta} H_{0}[Q(\xi) ; \xi \rightarrow r]=0 \quad (r > a),
\end{aligned}
\end{equation}
from which we can further derive the governing equation of $A(\xi)$ as
\begin{equation}
\begin{aligned}
H_{0}[\xi A ; \xi \rightarrow r]=&\frac{3-4 \eta}{4 \mu(1-\eta)} H_{0}\left[\xi^{-1} Q(\xi) ; \xi \rightarrow r\right] \quad (0\leq r\leq a),\\
H_{0}\left[\xi^{2} A ; \xi \rightarrow r\right]=&\frac{1-\eta}{\mu} H_{0}[Q(\xi) ; \xi \rightarrow r] \quad (r > a).
\end{aligned}
\label{governing_A_ii}
\end{equation}
However, it is not straightforward to explicitly obtain the solution $A(\xi)$ from \eqref{governing_A_ii}.
Instead, if we assume that $\xi^2 A(\xi)=\frac{1-\eta}{\mu}Q(\xi)+T(\xi)$, then the governing equation \eqref{governing_A_ii} can be further reduced to 
\begin{equation}
\begin{aligned}
&H_{0}\left[\xi^{-1} T(\xi) ; \xi \rightarrow r\right]=\frac{-(1-2 \eta)^{2}}{4 \mu(1-\eta)} Q(\xi) H_{0}\left[\xi^{-1} Q(\xi) ; \xi \rightarrow r\right], \quad (0\leq r\leq a), \\
&H_{0}[T(\xi) ; \xi \rightarrow r]=0, \quad (r > a).
\end{aligned}
\label{governing_T}
\end{equation}
Then, following \cite{sneddon1960elementary}, Eq. \eqref{governing_T} is satisfied if 
\begin{equation}
\begin{aligned}
T(\xi)=&\int_0^a \gamma(t) \cos(\xi t) \mathrm{d} t, \\
\gamma(t)=&\frac{(1-2 \eta)^{2}}{2 \pi \mu(1-\eta)}\left[\int_{0}^{\infty} \mathcal T_{ r}^{\prime}(r, 0) d r-\int_{0}^{t}\frac{t \mathcal T_{ r}^{\prime}(r, 0) }{\sqrt{t^{2}-r^{2}}}d r\right].
\end{aligned}
\label{solutionT_ii}
\end{equation}
Hence, by combining Eqs. \eqref{sigmaPz0_ii}, \eqref{AB_relationship_ii}, and \eqref{solutionT_ii}, the final solutions for $w_z^{\prime\prime}$ and $\sigma_{z z}^{\prime\prime}$ of the linear problem $(ii)$ in Eq. \eqref{EQsecond-2} are given by
\begin{equation}
\begin{aligned}
&\left.w_{z}^{\prime \prime}\right|_{z=0} =
 \begin{cases}
 0, & \quad (0\leq r\leq a),\\
\dfrac{2(1-\eta)}{1-2 \eta} \int_{0}^{a} \dfrac{\gamma(t)}{\sqrt{r^{2}-t^{2}}} \mathrm{~d} t -\dfrac{(1-2 \eta)}{2 \mu} \int_{r}^{\infty} \mathcal T_{ r}^{\prime}(\xi, 0) d \xi, & \quad ( r > a),
 \end{cases}
\\
&\left.\sigma_{z z}^{\prime \prime}\right|_{z=0}=
\begin{cases}
\dfrac{-2 \mu}{1-2 \eta}\left(\dfrac{\gamma(a)}{\sqrt{a^{2}-r^{2}}}-\int_{r}^{a} \dfrac{\gamma^{\prime}(t)}{\sqrt{t^{2}-r^{2}}} \mathrm{~d} t\right), & \quad (0\leq r\leq a),\\
 0, & \quad ( r > a).
 \end{cases}
\end{aligned}
\end{equation}

\subsubsection{Solving linear problem $(iii)$}
The linear problem $(iii)$ of Eq. \eqref{EQsecond-3} corresponds to the well-known linear contact problem with a flat indenter and zero body force. Hence, we can adopt the solution of \cite{sneddon1965relation}, which 
gives 
\begin{equation}
\begin{aligned}
&\left.w_{z}^{\prime \prime\prime}\right|_{z=0} =
 \begin{cases}
D_2 &\quad (0\leq r\leq a),\\
\dfrac{2  D_2}{\pi}\arcsin (a/r) & \quad ( r > a),
 \end{cases}
\\
&\left.\sigma_{z z}^{\prime \prime\prime}\right|_{z=0}=
\begin{cases}
-\dfrac{2 \mu D_2}{\pi (1-\eta)} \left(a^2-r^2\right)^{-1/2}& \quad (0\leq r\leq a),\\
 0, & \quad ( r > a).
 \end{cases}
\end{aligned}
\end{equation}

\subsection{Closing the second-order elastic problem}
In the proceeding subsection, we have derived the required second-order solutions at the contact surface $z=0$. By linear superposition, the final solutions of this second-order contact problem are the sums of these separate solutions. For the displacement $u_z$ and $t_{zz}$, these are given by
\begin{equation}
\begin{aligned}
u_z=&v_z+w_z^{\prime}+w_z^{\prime\prime}+w_z^{\prime\prime\prime}\\
t_{zz}=&\tau_{zz}+\tau_{zz}^{\prime}+\sigma_{zz}^{\prime}+\sigma_{zz}^{\prime\prime}+\sigma_{zz}^{\prime\prime\prime}
\end{aligned}.
\label{uz&tzz}
\end{equation}
In addition, the applied force $P$ is
\begin{equation}
P=-2 \pi \int_{0}^{a} r \left. t_{z z}\right|_{z=0} \mathrm{d} r.
\label{P}
\end{equation}

\section{Asymptotic solution for spherical indentation}
In this section, we focus on one of the most common indentation problems using a spherical indenter.
For the rigid spherical indenter with radius $R$, the profile function is 
\begin{equation}
f(r) =R-\sqrt{R^{2} -r^{2}}.
\end{equation}
\begin{figure}[h]
\begin{center}
\includegraphics[width=1\textwidth]{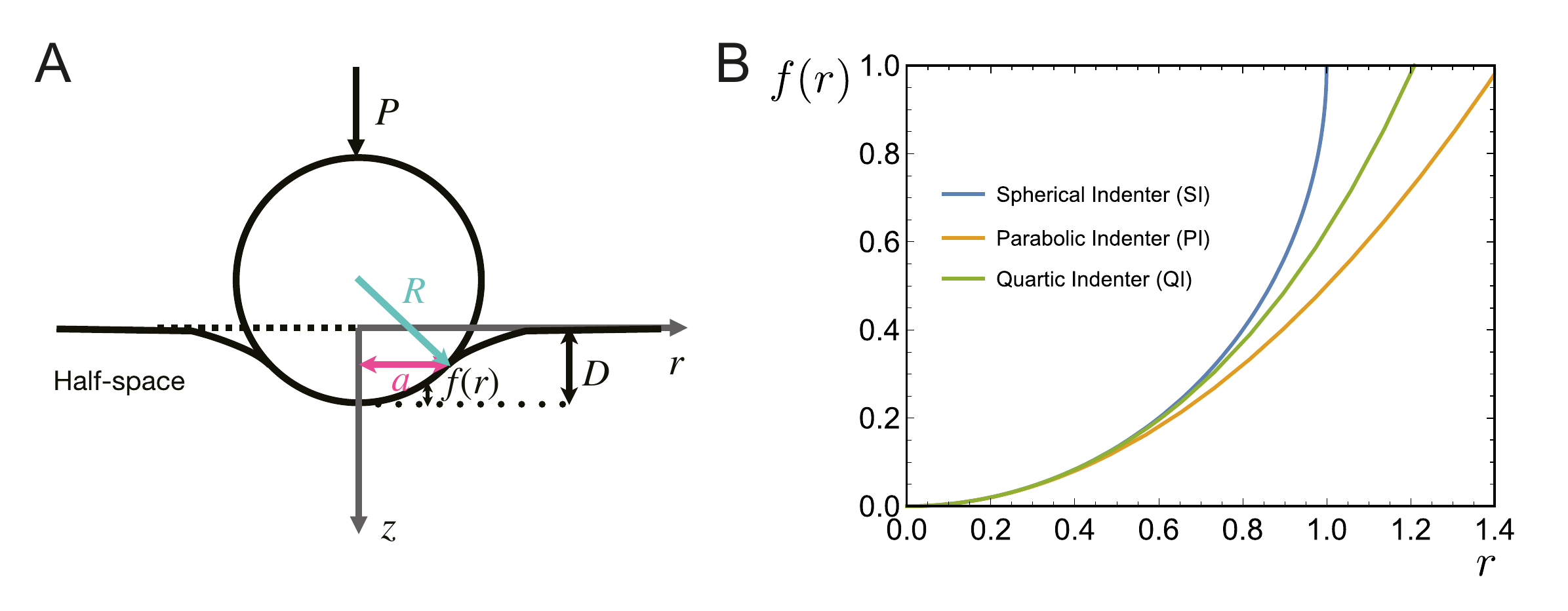}
\caption{(A) Diagram of the finite spherical indentation. (B) Profile functions for different indenters (R=1).}
\label{Spherical-Indenter_combined}
\end{center}
\end{figure}
\cite{sneddon1965relation} obtained the first-order analytical solution, but in this case, we have not been able to find the second-order analytical solution with this spherical function. The explicit integrals could not be found for Eq. \eqref{FGpsichi}$_{1,2}$. In the following subsections, we derive two asymptotic analytical solutions using parabolic and quartic surfaces. 

\subsection{Asymptotic solutions using a parabolic surface}
As a simplification of the indenter profile adopted by \cite{hertz1881contact},  we use an axisymmetric paraboloid to obtain an asymptotic analytical solution of spherical indentation up to the second-order in perturbation amplitude.
As shown in Figure \ref{Spherical-Indenter_combined}$B$, the spherical profile function $f(r)$ can be approximated by a parabolic function
\begin{equation}
f(r)=\frac{r^{2}}{2R} + O\left(\frac{r^4}{R^3}\right).
\label{fr_parabolic}
\end{equation}
Furthermore, given $\varepsilon=a/R$ and substituting \eqref{fr_parabolic} into \eqref{FGpsichi}$_4$, we obtain  
\begin{equation}
\chi(t)=\frac{2}{\pi}\left(D_1-\varepsilon \frac{t^2}{a}\right),
\end{equation}
by which, recalling that $\chi(a)=0$ is required, we find 
\begin{equation}
D_1=\varepsilon a.
\end{equation}
Hence, according to \eqref{FGpsichi}$_3$, we have
\begin{equation}
\psi(\xi)=\frac{2 \varepsilon \left(\sin(a \xi)-a \xi \cos(a \xi) \right)}{\pi a(1-\eta) \xi^3}.
\end{equation}
Based on \eqref{FGpsichi}$_1$ and \eqref{FGpsichi}$_2$,  $F_i(r,z)$ and $G_i(r,z)$ are given by
\begin{equation}
\begin{aligned}
F_{i}(r,z)=\int_{0}^{\infty} \xi^{i} \psi(\xi) \mathrm{e}^{-\xi z} J_{0}(\xi r) d \xi, \\
G_{i}(r,z)=\int_{0}^{\infty} \xi^{i} \psi(\xi) \mathrm{e}^{-\xi z} J_{1}(\xi r) d \xi.
\end{aligned}
\label{FiGi}
\end{equation}
At $z=0$, \eqref{FiGi} can be rewritten in the form of the Weber-Sonin-Schafheitlin integral \citep{korenev2002bessel}:
\begin{equation}
\begin{aligned}
F_{i}(r,0)=&\sqrt{\frac{2 a}{\pi}}\frac{\varepsilon}{ (1-\eta)}\int_0^\infty\frac{J_{3/2} (a \xi) J_0(r \xi)}{\xi^{3/2-i}}d\xi,\\
G_{i}(r,0)=&\sqrt{\frac{2 a}{\pi}}\frac{\varepsilon}{ (1-\eta)}\int_0^\infty\frac{J_{3/2} (a \xi) J_1(r \xi)}{\xi^{3/2-i}}d\xi.
\end{aligned}
\end{equation}
See Appendix A for the details.
Recalling Eqs. \eqref{uz&tzz}, for $r>a$, we have the displacement field
\begin{equation}
\begin{aligned}
u_{z} (r,0)=&\dfrac{\varepsilon}{ \pi a }\left(\left(r^2-2 a^2\right) \arcsin\left(\dfrac{a}{r}\right)-a \sqrt{r^2-a^2}\right)\\
&+w_z^{\prime}(r,0)+w_z^{\prime\prime}(r,0)+\dfrac{2 D_2}{\pi}\arcsin \left(\dfrac{a}{r}\right),
\end{aligned}
\end{equation}
where
\begin{equation}
\begin{aligned}
 w_{z}^{\prime}(r,0)=&-\frac{ 8 \varepsilon^{2} \sqrt{r^{2}-a^{2}}}{3 \pi^{2}} 
 - \frac{\left(3 a_{1}-4 a_{5}\right)(1-2 \eta)^{2}\varepsilon^{2}}{9 \pi^{2} a_{1}(1-\eta)} \frac{a^{4}}{r^{3}}\left( \frac{r\sqrt{r^{2}-a^{2}}}{a^{2}}+\right.\\
&\left. \ln \left(\frac{r+\sqrt{r^{2}-a^{2}}}{a}\right)\right)+
\frac{16 \varepsilon^{2}}{3 \pi^{3} a} \int_{a}^{r}\left[2 a\left(t^{2}-a^{2}\right) I_{1}+\left(3 a^{2}-2 t^{2}\right) I_{2}\right.\\
&\left.-\left(t^{2}-a^{2}\right) \ln \left(t^{2}-a^{2}\right) \arcsin \left(\frac{a}{t}\right)\right] \frac{\mathrm{d} t}{\sqrt{r^{2}-t^{2}}},
\end{aligned}
\end{equation}
\begin{equation}
\begin{aligned}
w_{z}^{\prime \prime }(r,0)=&\frac{(1-2\eta )^{2} \varepsilon ^{2}}{3\pi ^{2} (1-\eta )}\left(\frac{a^2\sqrt{r^{2} -a^{2}}}{r^{2}} +a\arcsin\left(\frac{a}{r}\right)\right) -\frac{2(1-2\eta )^{2} \varepsilon ^{2}a}{3\pi ^{2} \left(1-\eta \right)} I_4 \\
        &-\frac{(1-2\eta )^{2} \varepsilon ^{2}}{18\pi ^{2} a^{2} (1-\eta )^{2}}\left\{3a^{3} \left(1-2\eta \right)\left(1-2\ln a\right)\arcsin\left(\frac{a}{r}\right)\right.\\
       &\left.+\left[ \left(2\eta +5\right)a^{2}\sqrt{r^{2} -a^{2}} +6a^3\left(1-2\eta \right)I_3\right.\right.\\
       &\left.\left.+\sqrt{r^{2} -a^{2}}\left( \left(8\eta +2\right)a^{2}-\left(2\eta +5\right)r^{2}\right)\ln\left(\frac{r^{2}}{r^{2} -a^{2}}\right)\right]\right\},
\end{aligned}
\end{equation}
and $I_1$, $I_2$, $I_3$, and $I_4$ are listed in Appendix C.
Similarly, for $r<a$, we have the stress component
\begin{equation}
\begin{aligned}
t_{zz}(r,0)=&\frac{8 a_1 \varepsilon \sqrt{a^2-r^2}}{\pi a (1-\eta )}+\tau_{zz}^\prime(r,0)
+\sigma_{zz}^\prime(r,0)+\sigma_{zz}^{\prime\prime}(r,0)\\
&-\frac{2 \mu D_2}{(1-\eta ) \pi \sqrt{a^2-r^2}},
\end{aligned}
\end{equation}
where
\begin{equation}
\begin{aligned}
\tau_{zz}^\prime(r,0)=&\frac{2 \varepsilon ^2}{9 a^2 \left(1- \eta \right)^2} \left[\frac{36 \left(1-2 \eta \right)^2 \left(a_1+4a_2+12a_3+48 a_4\right) \left(a^2-r^2\right) }{\pi ^2}\right.\\
&\left.-\frac{24 \left(1-2 \eta \right)^2 \left(a_1+2 a_2-2 a_3-2 a_5\right)\sqrt{a^2-r^2} \left(a^3-\left(a^2-r^2\right)^{3/2}\right) }{\pi ^2 r^2}\right.\\
&\left.+\frac{8 (1-2 \eta )^2 \left(a_1+2 a_2-2 a_3-2 a_5\right) \left(a^3-\left(a^2-r^2\right)^{3/2}\right)^2 }{\pi ^2 r^4}\right.\\
&\left.+9 \left(1-\eta \right)^2 r^2 \left(a_1+4 a_2\right)\right],
\end{aligned}
\end{equation}
\begin{equation}
\begin{aligned}
\sigma_{zz}^\prime\left(r,0\right)=
&\frac{4 a^3 \left(1-2 \eta \right)^2 \varepsilon ^2 \left(3 a_1-4 a_5\right)}{9 \pi ^2 \left(1-\eta\right)^2 \left(2 a^3+2 a^2 \sqrt{a^2-r^2}-r^2 \sqrt{a^2-r^2}-2 a r^2\right)}\\
&+\frac{16 \mu \varepsilon ^2 I_7}{\pi ^3 a \left(1-\eta \right)}+\frac{8 a \mu \varepsilon ^2 \left(\ln 2-1\right)}{3 \pi ^2 \left(1-\eta\right) \sqrt{a^2-r^2}}+\frac{4 \mu \varepsilon ^2}{\pi ^2 \left(1-\eta \right)}\\
&+\frac{4 \varepsilon ^2}{9 \pi ^2 \left(1-\eta \right)^2}\left[2 a_1 \left(\eta -2\right) \left(4 \eta -5\right)-8 a_2 \left(1-2 \eta \right)^2+9 \pi ^2 a_2 \left(\eta -1\right)+8 a_3 \left(1-2 \eta \right)^2\right]\\
&-\frac{48 \left(1-2 \eta \right) \varepsilon ^2}{9 \pi ^2 \left(1- \eta \right)^2} \left[4 \eta \left(a_1+4 a_2-2 a_3-a_5\right)-5 a_1+4 \left(a_3-2 a_2+a_5\right)\right]   \ln\left(1+\sqrt{1-\frac{r}{a}^2}\right)  \\
&+\frac{8 \left(1-2 \eta \right)^2 \varepsilon ^2 \left(a_1-4 a_2+4 a_3\right)\left(2 a^4-a r^2 \sqrt{a^2-r^2}-2 a^3 \sqrt{a^2-r^2}\right) }{9 \pi ^2 r^4 \left(1-\eta \right)^2 }\\
&+\frac{2 \varepsilon ^2 \left[9 \pi ^2 a_1 \left(\eta -1\right)-4 a_1 \left(28 \eta ^2+8 \eta -11\right)\right]}{9 \pi ^2 \left(1-\eta \right)^2}\\
&+\frac{32 \left(1-2 \eta \right) \varepsilon ^2 \left[a_3 \left(68 \eta -43\right)+108 a_4 \left(2 \eta -1\right)+3 a_5 \left(\eta -1\right)\right]}{9 \pi ^2 \left(1-\eta \right)^2}\\
&-\frac{4 a_2 \varepsilon ^2 \left[9 \pi ^2 \left(\eta -1\right)-8 \left(4 \eta ^2-40 \eta +19\right)\right]}{9 \pi ^2 \left(1-\eta \right)^2}+\frac{2 a_1 \varepsilon ^2 r^2\left[80 \eta ^2+64 \eta -9 \pi ^2 \left(\eta -1\right)-52\right]}{9 \pi ^2 a^2 \left(1-\eta \right)^2}\\
&+\frac{32 \left(2 \eta -1\right)\varepsilon ^2 r^2 \left[a_3 \left(64 \eta -41\right)+108 a_4 \left(2 \eta -1\right)+3 a_5 \left(\eta -1\right)\right]}{9 \pi ^2 a^2 \left(1-\eta \right)^2}\\
&+\frac{4 a_2 \varepsilon ^2 r^2 \left[8 \left(4 \eta ^2+32 \eta -17\right)+9 \pi ^2 \left(\eta -1\right)\right]}{9 \pi ^2 a^2 \left(1-\eta \right)^2},
\end{aligned}
\end{equation}
\begin{equation}
\begin{aligned}
\sigma_{zz}^{\prime\prime}\left(r,0\right)=&\frac{2 a_1 (2 \eta -1) \varepsilon ^2}{9 \pi ^2 a (1-\eta)^2 \sqrt{a^2-r^2}} \left\{a^2 \left[\eta \left(22-12 \ln 2\right)+37+6\ln 2\right]\right.\\
&\left.+9 a \sqrt{a^2-r^2} \left[\left(2 \eta +5\right) I_6-2 I_5 \right]-9 \left(2 \eta +5\right) r^2\right\},
\end{aligned}
\end{equation}
and $I_{5}$, $I_{6}$, $I_{7}$, are listed in Appendix C.
To avoid the singularity of $t_{zz}$ induced by the terms with $\left(a^2-r^2\right)^{-1/2}$at $r=a$ , $D_2$ can be chosen as
\begin{equation}
\begin{aligned}
D_2=\frac{\varepsilon D_1 \left(2 \eta -1\right)^2 }{9 \pi a_1 \left(\eta -1\right)}\left(3 a_1\ln 2-2 a_1+4 a_5\right)
-\frac{\varepsilon D_1 \left(2 \eta -1\right)}{3 \pi \left(\eta -1\right)}+\frac{4 \varepsilon D_1 \left(\ln 2-1\right)}{3 \pi }.
\end{aligned}
\end{equation}
Hence, the total indentation depth is given by
\begin{equation}
\begin{aligned}
D=& D_1+ \frac{\varepsilon D_1 \left(2 \eta -1\right)^2 }{9 \pi a_1 \left(\eta -1\right)}\left(3 a_1\ln 2-2 a_1+4 a_5\right)
-\frac{\varepsilon D_1 \left(2 \eta -1\right)}{3 \pi \left(\eta -1\right)}\\
&+\frac{4 \varepsilon D_1 \left(\ln 2-1\right)}{3 \pi }.
\end{aligned}
\end{equation}
Based on Eq. \eqref{P}, the total force $P$ is given by 
\begin{equation}
\begin{aligned}
P=&\frac{16 a^2 a_1 \varepsilon }{3 \left(\eta -1\right)}-\frac{a^2 \varepsilon ^2 }{9 \pi (1-\eta )^2}\left\{9 \pi ^2 a_1 \left(1-\eta \right)^2+12 a_1 \eta \left(10 \eta -37\right)+174 a_1\right.\\
&\left.- \left(1-2 \eta\right) \left[ 288 \left(a_3 - 2 a_2\right) \left(1-\eta \right)- 18 a_2\pi ^2\left(1-\eta\right)- 16a_5 \left(\eta -5\right)\right]\right\}.
\end{aligned}
\end{equation}
For incompressible materials $\eta=1/2$, and the Lam\'e constants behave as
\begin{equation}
\begin{aligned}
\lambda =\frac{2\mu \eta }{1-2\eta } \rightarrow \infty, \quad
 \mu =\frac{E}{2( 1+\eta )} \rightarrow \frac{E}{3},
\end{aligned}
\end{equation}
which indicates that
\begin{equation}
\begin{aligned}
a_{1} =-\frac{\mu }{2} =-E/6, \quad a_{2} =\frac{\lambda +2\mu }{8}= \mathcal{O} (\lambda /\mu )\rightarrow \infty.
\end{aligned}
\label{a1a2}
\end{equation}
Moreover, according to \cite{destrade2010third}, for incompressible material, we have
\begin{equation}
\begin{aligned}
&a_{3} =\mathcal{O} (\lambda /\mu ),\quad
a_{4} =\mathcal{O}\left( \lambda ^{2} /\mu ^{2}\right), \quad
a_{5} =\mathcal{O}(\mu), \quad
\\
&a_{3}-2a_{2}=a_1-a_5=\mathcal{O}(\mu).
\end{aligned}
\label{a3a4a5}
\end{equation}
Note that the material constants $a_3, a_4$, and $a_5$ used in \cite{giannakopoulos2007spherical} do not satisfy these constraints, since $a_4$ was wrongly set to be of order $\mathcal{O} (\lambda /\mu )$ and $a_3-2a_2 \neq a_1-a_5$.

Next, according to \eqref{a1a2} and \eqref{a3a4a5}, for incompressible materials, the corresponding results reduce to
\begin{equation}
\begin{aligned}
u_{z} (r,0)=&\dfrac{\varepsilon}{ \pi a }\left(\left(r^2-2 a^2\right) \arcsin\left(\dfrac{a}{r}\right)-a \sqrt{r^2-a^2}\right)\\
&-\frac{ 8 \varepsilon^{2} \sqrt{r^{2}-a^{2}}}{3 \pi^{2}} +\dfrac{2 D_2}{\pi}\arcsin \left(\dfrac{a}{r}\right)\\
&+\frac{16 \varepsilon^{2}}{3 \pi^{3} a} \int_{a}^{r}\left[2 a\left(t^{2}-a^{2}\right) I_{1}+\left(3 a^{2}-2 t^{2}\right) I_{2}\right.\\
&\left.-\left(t^{2}-a^{2}\right) \ln \left(t^{2}-a^{2}\right) \arcsin \left(\frac{a}{t}\right)\right] \frac{\mathrm{d} t}{\sqrt{r^{2}-t^{2}}},~ (r> a)\\
t_{zz}(r,0)=&2 \mu \varepsilon ^2 \left(\frac{a^2-r^2}{a^2}+\frac{16 I_9}{\pi ^3 a}\right)-\frac{8 \mu \varepsilon \sqrt{a^2-r^2}}{\pi a}, ~(0\leq r\leq a)\\
D=&\varepsilon a+\frac{4 \varepsilon^2 a \left(\ln 2-1\right)}{3 \pi },\\
P=&\frac{16}{3} \mu \varepsilon a^{2}-\frac{4 \mu \varepsilon ^{2} a^{2} }{\pi }.
\end{aligned}
\label{PD2}
\end{equation}
It is worth noting that the analytical solutions \eqref{PD2} correct those given by both \cite{giannakopoulos2007spherical} and \cite{sabin1983contact}.
\cite{giannakopoulos2007spherical} used inappropriate material constants which do not satisfy the incompressibility constraints. 
In addition, we found that \cite{giannakopoulos2007spherical} used the same expression for the applied force $P$ as \cite{sabin1983contact}. 
Furthermore, it has been found \citep{liu2010nonlinear,zhang2014spherical} that, compared to the numerical simulation results, the applied forces predicted by both \cite{giannakopoulos2007spherical} and \cite{sabin1983contact} are significantly overestimated. 

\subsection{Asymptotic solutions using quartic surface}
As shown in Figure \ref{Spherical-Indenter_combined}$B$, the parabolic surface is not sufficiently accurate to approximate the spherical indenter as the indentation depth increases.
Alternatively, following \cite{liu2010nonlinear}, we can further expand the profile function $f(r)$ up to quartic surface $O(r^4/R^3)$, which gives 
\begin{equation}
f(r)=\frac{r^{2}}{2R}+\frac{r^4}{8 R^3} + O\left(\frac{r^6}{R^5}\right).
\label{fr_quartic}
\end{equation}
Substituting \eqref{fr_quartic} into \eqref{FGpsichi}$_4$, we obtain  
\begin{equation}
\chi(t)=\frac{2 D_1}{\pi }-\frac{2 t \left(3 R^2 t+t^3\right)}{3 \pi  R^3}.
\end{equation}
Recalling that the boundary conditions require $\chi(a)=0$, we get 
\begin{equation}
D_1=\varepsilon a \left(1+\frac{\varepsilon ^2}{3}\right).
\label{D14}
\end{equation}
Note that following earlier authors, we include $O(\varepsilon^3)$ terms in  $D$, instead of pursuing a formal expansion of $D$ to $O(\varepsilon^3)$ i.e. $D=\varepsilon D_1 + \varepsilon^2 D_2 +\varepsilon^3 D_3+...$.
Next, according to \eqref{FGpsichi}$_3$, we have
\begin{equation}
\begin{aligned}
\psi(\xi)=&\frac{2 \varepsilon  \left(2 \varepsilon ^2+3\right)}{3 \pi  a \left(1-\eta \right)}\frac{ \sin (a \xi )- a \xi  \cos (a \xi )}{\xi ^3 }\\
&+\frac{8 \varepsilon ^3}{3 \pi  a^3 \left(1-\eta \right)}\frac{a^2 \xi ^2 \sin (a \xi )-3 \sin (a \xi )+3 a \xi  \cos (a \xi )}{\xi ^5 }.
\end{aligned}
\label{psi_q}
\end{equation}
Based on \eqref{psi_q},  $F_{i}(r,z)$ and $G_{i}(r,z)$ in terms of  \eqref{FiGi} at $z=0$ can be rewritten as
\begin{equation}
\begin{aligned}
F_{i}(r,0)=&\sqrt{\frac{2 a }{\pi}} \frac{ \varepsilon  \left(3+2 \varepsilon ^2\right)}{3 \left(1-\eta \right)}\int_0^\infty\frac{J_{3/2} (a \xi) J_0(r \xi)}{\xi^{3/2-i}}d\xi\\
&-\sqrt{\frac{2}{\pi a}} \frac{ 4\varepsilon ^3}{3 \left(1-\eta \right)}\int_0^\infty\frac{J_{5/2} (a \xi) J_0(r \xi)}{\xi^{5/2-i}}d\xi,
\\
G_{i}(r,0)=&\sqrt{\frac{2 a }{\pi}} \frac{ \varepsilon  \left(3+2 \varepsilon ^2\right)}{3 \left(1-\eta \right)}\int_0^\infty\frac{J_{3/2} (a \xi) J_1(r \xi)}{\xi^{3/2-i}}d\xi\\
&-\sqrt{\frac{2}{\pi a}} \frac{ 4\varepsilon ^3}{3 \left(1-\eta \right)}\int_0^\infty\frac{J_{5/2} (a \xi) J_1(r \xi)}{\xi^{5/2-i}}d\xi.
\end{aligned}
\end{equation}
See Appendix B for the details about $F_{i}(r,0)$ and $G_{i}(r,0)$, $i=0,~1,~2$.

Next, to simplify the calculation in this case, we shall only provide the solution for incompressible materials. 
For $0\leq r\leq a$, the stress component $t_{zz}$ is given by 
\begin{equation}
\begin{aligned}
t_{zz}(r,0)=&\frac{8 a_1 \varepsilon  \sqrt{a^2-r^2} \left(a^2 \left(2 \varepsilon ^2+9\right)+4 r^2 \varepsilon ^2\right)}{9 \pi  a^3 (1-\eta )}+\frac{r^2 (a_1+4 a_2) \left(2 a^2 \varepsilon +r^2 \varepsilon ^3\right)^2}{2 a^6}\\
&+\frac{128 a_4 (1-2 \eta )^2 \varepsilon ^2 \left(a^2-r^2\right) \left(a^2 \left(2 \varepsilon ^2+9\right)+4 r^2 \varepsilon ^2\right)^2}{27 \pi ^2 a^6 (\eta -1)^2}\\
&+\sigma_{zz}^\prime(r,0)+0-\frac{2\mu D_2}{(1-\eta ) \pi  \sqrt{a^2-r^2}},
\end{aligned}
\label{tzz41}
\end{equation}
where
\begin{equation}
\begin{aligned}
\sigma_{zz}^\prime(r,0)=&I_8+\frac{\varepsilon ^2}{945 \pi ^2} \left\{\frac{70 r^6 \varepsilon ^4 \left[9 \pi ^2 (a_1-3 a_2)+4096 a_4 (1-2 \eta )^2\right]}{a^6}\right.\\
&\left.+\frac{315 r^4 \varepsilon ^2 \left[3 \pi ^2 (3 a_1-8 a_2)+4096 a_4 (1-2 \eta )^2\right]}{a^4}\right.\\
&\left.-\frac{420 r^2 \left[128 a_4 (1-2 \eta )^2 \left(4 \varepsilon ^4+12 \varepsilon ^2-27\right)-9 \pi ^2 (a_1-2 a_2)\right]}{a^2}\right.\\
&\left.+\frac{24 a a_1 \left[\varepsilon ^4 (96 \ln2-38)+21 \varepsilon ^2 (3+4\ln2)-420 (\ln2-1)\right]}{\sqrt{a^2-r^2}}\right.\\
&\left.-35 \left[9 \pi ^2 a_1 \left(2 \varepsilon ^4+9 \varepsilon ^2+12\right)+512 a_4 (1-2 \eta )^2 \left(2 \varepsilon ^2+9\right)^2\right]\right\},
\end{aligned}
\end{equation}
and $I_8$ is given in Appendix C. 
To avoid the singularity of $t_{zz}$ at $r=a$, $D_2$ can be chosen as
\begin{equation}
\begin{aligned}
D_2= \frac{ 4 \varepsilon^2 a (\ln2-1)}{3 \pi }-\frac{ \varepsilon^4 a (4\ln2+3)}{15 \pi }-\frac{  \varepsilon^6a (96 \ln2-38)}{315 \pi },
\end{aligned}
\label{D24}
\end{equation}
and, therefore, we obtain the total indentation depth as
\begin{equation}
\begin{aligned}
D=a \varepsilon+\frac{4 \varepsilon ^2a (\ln2-1) }{3 \pi }+\frac{a \varepsilon ^3}{3}-\frac{ \varepsilon ^4 a (4\ln2+3)}{15 \pi }-\frac{  \varepsilon ^6a (96 \ln2-38)}{315 \pi }.
\end{aligned}
\label{D4}
\end{equation}
In addition, according to \eqref{D24}, and \eqref{a1a2}, \eqref{a3a4a5} for incompressible materials,  $t_{zz}(r,0)$ in \eqref{tzz41} can be further simplified to
\begin{equation}
\begin{aligned}
t_{zz}(r,0)=&I_8-\frac{8 \mu  \varepsilon  \sqrt{a^2-r^2}}{\pi  a}+\frac{2 \mu  \varepsilon ^2 \left(a^2-r^2\right)}{a^2}-\frac{16 \mu  \varepsilon ^3 \left(a^2+2 r^2\right) \sqrt{a^2-r^2}}{9 \pi  a^3 }\\
&+\frac{3 \mu  \varepsilon ^4 \left(a^4-r^4\right)}{2 a^4 }+\frac{\mu  \varepsilon ^6 \left(a^6-r^6\right)}{3 a^6 }.
\end{aligned}
\end{equation}
Finally, based on Eq. \eqref{P}, the total force $P$ is given by 
\begin{equation}
\begin{aligned}
P=&\frac{16}{3} \mu a^2   \varepsilon-\frac{4 }{\pi }\mu a^2   \varepsilon ^2+\frac{32}{15} \mu a^2   \varepsilon ^3-\frac{16 \ln2}{3 \pi } \mu  a^2  \varepsilon ^4 -\frac{ 96 \ln2-28}{45 \pi } \mu a^2  \varepsilon ^6,
\end{aligned}
\label{P4}
\end{equation}
where the higher-order terms are generated by the quartic profile function \eqref{fr_quartic} and needed to avoid the singularity of $\sigma_{zz}^\prime(r,0)$ that would otherwise appear at $r=a$. Equation \eqref{psi_q}--\eqref{P4} contain terms that are $O(\varepsilon^3)$ or higher, beyond the $O(\varepsilon^2)$ expansion of the deformation field. Retaining these extra terms that arise from the indenter shapes greatly improves the agreement with the FE simulations.

\section{Results and discussion}
Table \ref{table1} shows a summary of both first- and second-order indentation models and their analytical solutions for the indentation force and displacement.
The first-order indentation models include the Hertz model, Liu's model \citep{liu2010nonlinear}, and Sneddon's model \citep{sneddon1965relation}, that are derived using the parabolic, quartic, and spherical profile functions, respectively. 
The second-order indentation models, analytical parabolic and analytical quartic are derived using the parabolic and quartic profile functions.
In the following subsection, we verify these indentation models by comparison with finite element (FE) simulations.

\begin{table}
\caption{Summary of the indentation models for incompressible materials}
\label{table1}
\center
\renewcommand{\arraystretch}{2.5}
\hspace*{0cm}\begin{tabular}{ | m{4cm} | m{7cm}| m{4cm} | } 
 \hline
Name of model \newline (Theoretical method used)& Analytical solutions of the force and displacement & Profile of the indenter $f(r)$ used in calculation \\ 
 \hline
  
Sneddon model \newline(First-order elasticity)
& $D_{S} =\dfrac{1}{2} a\ \ln\left(\dfrac{R+a}{R-a}\right)$ \newline $P_{S} =2\mu \left[\left( a^{2} +R^{2}\right)\ln\left(\dfrac{R+a}{R-a}\right) -2aR\right]$ 
& $f( r) =R-\sqrt{R^{2} -r^{2}}$ \\ 
 \hline
  
Hertz model \newline(First-order elasticity) 
& $D_{H} =\varepsilon a$ \newline $P_{H} =\dfrac{16}{3} a^{2} \varepsilon \mu $ 
& $ f( r) =\dfrac{r^{2}}{2R}$ \\ 
 \hline
  
 Liu's model \newline(First-order elasticity) 
 & $D_{L} =\varepsilon a+\dfrac{\varepsilon ^{3} a}{3}$ \newline $P_{L} =\dfrac{16}{3} a^{2} \varepsilon \mu +\dfrac{32}{15} a^{2} \varepsilon ^{3} \mu $ \
 & $ f(r)=\dfrac{r^{2}}{2R}+\dfrac{r^4}{8 R^3} $ \\ 
 \hline
  
 Analytical parabolic \newline(Second-order elasticity) 
 & $D _{P} =\varepsilon a-\dfrac{4a \varepsilon ^{2}}{3\pi }( 1-\ln 2)$ 
\newline $P_{P} =\dfrac{16}{3} a^{2} \varepsilon \mu -\dfrac{4a^{2} \varepsilon ^{2} \mu }{\pi } $ 
& $ f( r) =\dfrac{r^{2}}{2R}$ \\
 \hline
  
 Analytical quartic \newline(Second-order elasticity) 
& $D _{Q} =\varepsilon a+\dfrac{\varepsilon ^{3} a}{3} -\dfrac{4a \varepsilon ^{2}}{3\pi }( 1-\ln 2) \quad \newline~~~~~~~~~~~
-\dfrac{a \varepsilon ^{4}}{15\pi }( 3+4\ln 2)
+\dfrac{a \varepsilon ^{6}}{315\pi }( 38-96\ln 2)$ 
\newline $P_{Q} =\dfrac{16}{3} a^{2} \varepsilon \mu -\dfrac{4a^{2} \varepsilon ^{2} \mu }{\pi } +\dfrac{32}{15} a^{2} \varepsilon ^{3} \mu \newline~~~~~~~~~~~ -\dfrac{16a^{2} \varepsilon ^{4} \mu \ln 2}{3\pi }+\dfrac{4a^{2} \varepsilon ^{6} \mu }{45\pi }( 7-24\ln 2)$ 
& $ f(r)=\dfrac{r^{2}}{2R}+\dfrac{r^4}{8 R^3} $ \\
 \hline
\end{tabular}
\renewcommand{\arraystretch}{2.5}
\center
\end{table}

\subsection{Finite element simulations}
ABAQUS (2017) \citep{Abaqus2017} is used to simulate the indentation problem with nonlinear deformation.
To simplify the calculation, we establish axisymmetric models for both the indenter and substrate. 
The indenter is assumed to be a rigid body with radius $R = 3 \text{mm}$, while the half-space substrate is assumed to be an incompressible neo-Hookean solid and mimicked by a finite cylinder with appropriate scale and boundary conditions. For the incompressible neo-Hookean solid, the energy function is
\begin{equation}
W_{NH}=C_{10}(\bar I_1-3),
\end{equation}
where $C_{10}$ is the material constant, $\bar I_1=\mathrm{tr}\left(\boldsymbol{F}^T\boldsymbol{F}\right)$ is the first invariant, and $\boldsymbol{F}$ is the deformation gradient tensor.
Up to third-order, the energy function of an incompressible neo-Hookean solid can be expanded as
\begin{equation}
W_{NH}^*=-C_{10}J_{2}+\frac{C_{10}}{2}J_1^2-C_{10}J_{1}J_{2}+\frac{C_{10}}{3}J_1^3+C_{10}J_{3},
\end{equation}
so $C_{10}=-a_1$.
Here, we specify the nonlinear material constant $C_{10}=\mu/2=0.15 \text{MPa}$. The maximum finite indentation depth is set to be the same as the indenter radius $R$, that is, $D/R = 1.0$. 
In addition, we use 2-node linear axisymmetric rigid elements to discretise the indenter, and 4-node axisymmetric reduced integration hybrid elements (CAX4RH) and some 3-node bilinear axisymmetric hybrid elements (CAX3H) to discretise the half-space body. 

\begin{figure}[h]
\begin{center}
\hspace*{0cm} 
\includegraphics[width=1\textwidth]{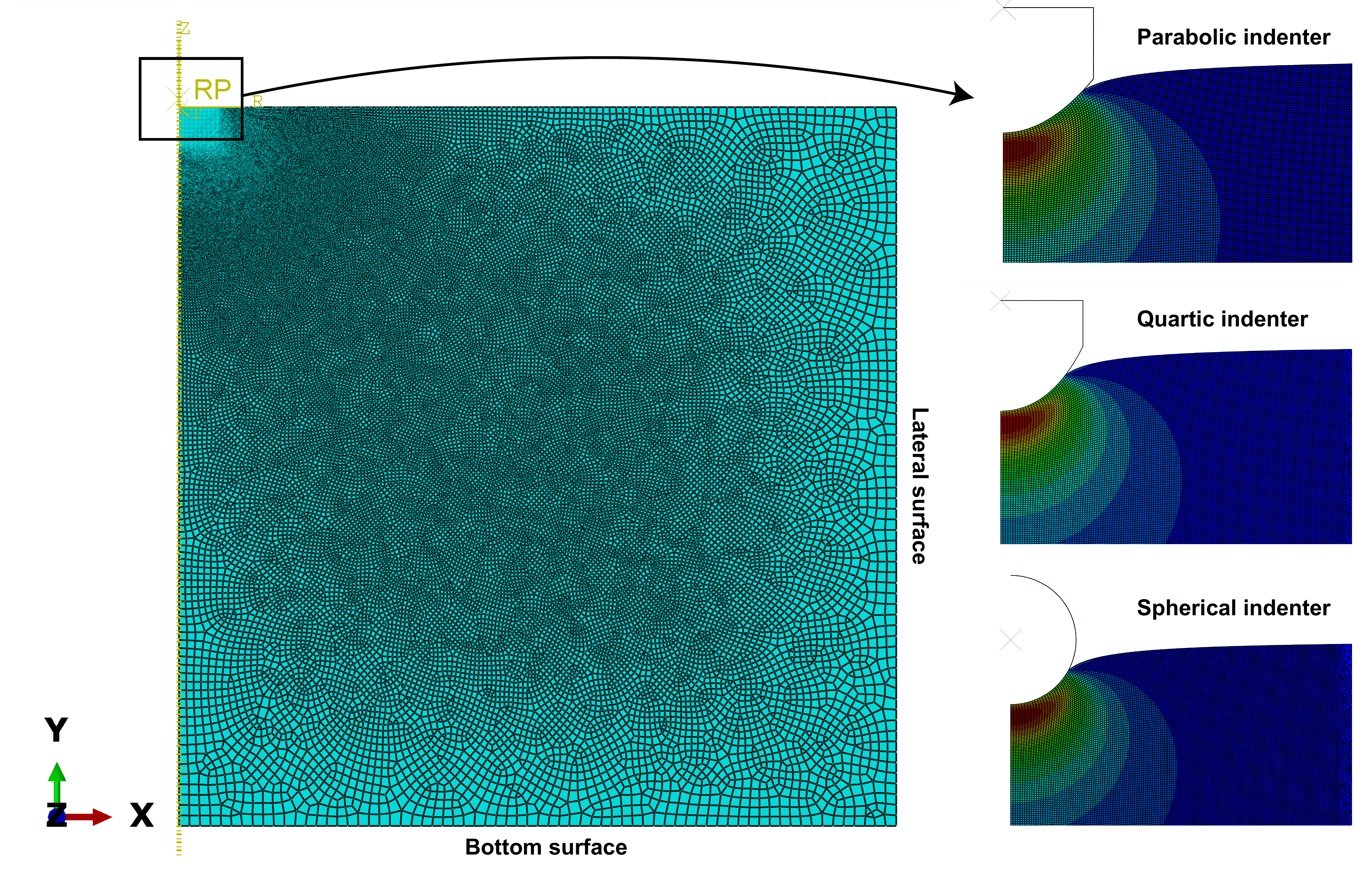}
\caption{The FE models established in ABAQUS, including parabolic, quartic, and spherical indentations ($R=3\text{mm}$), showing the mesh and the indenter shapes.  Axisymmetric models are established for simplifying the calculation. The maximum finite indentation depth is set to be the same as the indenter radius $R$, that is, $D/R = 1.0$.  2-node linear axisymmetric rigid elements are used to discretise the indenter, and 4-node axisymmetric reduced integration hybrid elements (CAX4RH) and some 3-node bilinear axisymmetric hybrid elements (CAX3H) are used to discretise the half-space body.}
\label{FE-Models}
\end{center}
\end{figure}

 

Compared to an infinite half-space, using a finite scale cylinder requires that we should specify additional information, including its size and the boundary conditions.
Therefore, we should very carefully consider these two influential factors. 
With reference to Appendix D, we studied their impact by setting control groups and then determining an appropriate scale and boundary conditions.  
The results show that establishing the FE model for a cylinder with radius and height of $270\text{mm}$ and boundary condition $u_{Bz}=0$ (i.e. the displacement of the bottom surface along the $z$-direction is constrained) to represent the half-space substrate is appropriate.

\subsubsection*{Influence of the indenter shapes}
\begin{figure}[h]
\begin{center}
\includegraphics[width=1\textwidth]{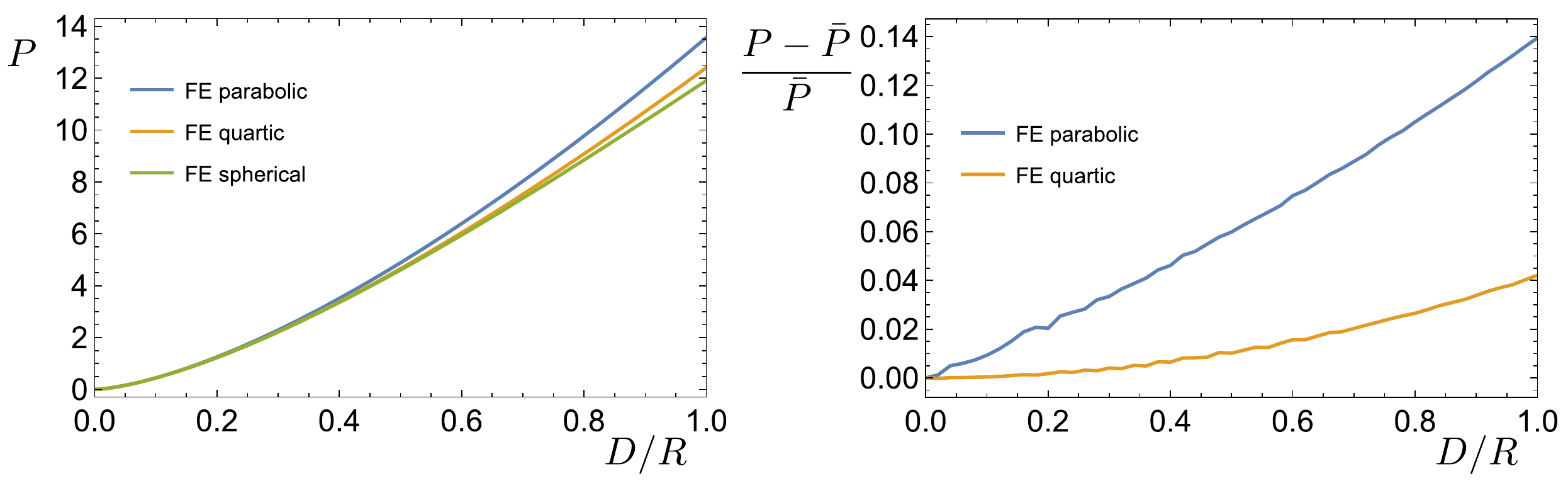}
\caption{Comparison of FE simulation results between parabolic, quartic, and spherical indenters. (A) The corresponding force-displacement curves; (B) the percentage difference of FE parabolic and quartic indentation to the FE spherical indentation, where $\bar{P}$ is the force by FE spherical indentation.}
\label{FEM-Different-Indenters}
\end{center}
\end{figure}

In Figure \ref{FEM-Different-Indenters}$A$, we display the comparison of the force-displacement curves obtained from FE simulations with different indenter shapes. 
Figure \ref{FEM-Different-Indenters}$B$ shows the percentage differences for the parabolic and quartic indentations relative to the spherical indentation, where both the parabolic and quartic indenters provide an overestimate of the applied force. 
This is explained by Figure \ref{Spherical-Indenter_combined}$B$, where we see that, at the same indentation depth, both the parabolic and quartic indenters have larger contact areas.
According to Figure \ref{FEM-Different-Indenters}$B$, using the parabolic indenter would make a 4\% difference when the indentation depth $D/R\approx0.4$, while using the quartic indenter would only exhibit the same difference when $D/R\approx 1.0$.
Moreover, at the maximum indentation depth of $D/R=1.0$, the difference in using the quartic indenter is three times smaller than that using the parabolic indenter.
Therefore, the quartic indenter is indeed a much better approximation of the spherical indenter than the parabolic indenter, as expected.

\subsection{Verification of indentation models}
In this section, we use FE simulations, including parabolic, quartic, and spherical indenter profiles as benchmark results, and verify the indentation models in Table \ref{table1} by comparing them to their corresponding FE results.

\subsubsection*{Parabolic indentation}
\begin{figure}[h]
\begin{center}
\includegraphics[width=1\textwidth]{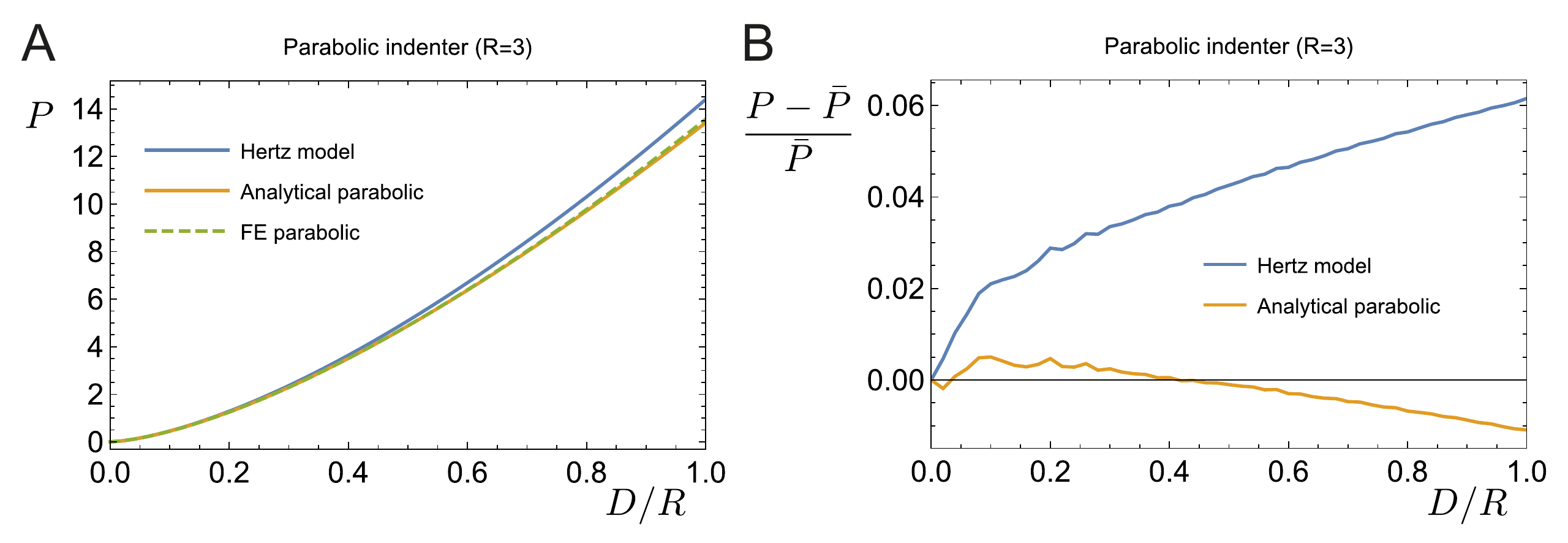}
\caption{Comparisons among the first-order Hertz model, the second-order analytical parabolic solution, and the FE simulation result of the parabolic indentation. (A) The corresponding force-displacement curves; (B) the percentage difference of Hertz model and second-order analytical parabolic solution to the FE parabolic indentation, where $\bar{P}$ is the force by FE parabolic indentation.}
\label{PIPD-Error}
\end{center}
\end{figure}

Figure \ref{PIPD-Error}$A$ shows the force-displacement curves from the first-order Hertz model, the second-order analytical parabolic solution \eqref{PD2}, and the FE computation of parabolic indentation.
Figure \ref{PIPD-Error}$B$ displays the differences between the Hertz model and the second-order analytical parabolic solution compared to the FE computation with a parabolic indentation. 
As shown in Figure \ref{PIPD-Error}$A$, the classical Hertz model overestimates the external force. In contrast, the second-order analytical parabolic solution exhibits very well-matched results over the entire indentation process. In particular, according to Figure \ref{PIPD-Error}$B$, the percentage difference of the analytical parabolic solution is only slightly more than 1\% at the maximal indentation depth of $D/R=1.0$.
As a comparison, the percentage difference of the first-order solution Hertz model is over 1\% at the indentation depth $D/R\approx0.05$, and over 6\% at the maximal indentation depth $D/R=1.0$.

\subsubsection*{Quartic indentation}
\begin{figure}[h]
\begin{center}
\includegraphics[width=1\textwidth]{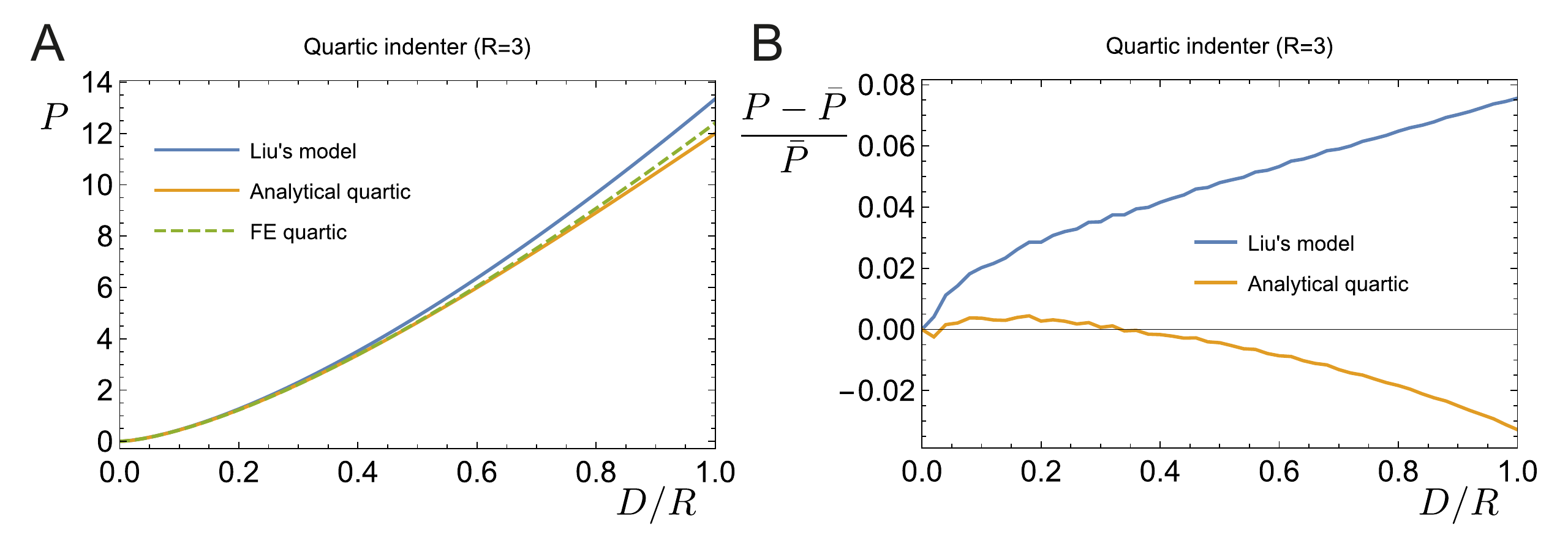}
\caption{Comparisons among the first-order Liu's model, the second-order analytical quartic solution, and the FE quartic indentation. (A) The corresponding force-displacement curves; (B) the percentage difference of Liu's model and second-order analytical quartic solution to the FE quartic indentation, where $\bar{P}$ is the force by FE quartic indentation.}
\label{FIPD-Error}
\end{center}
\end{figure}

Similarly, we show the force-displacement curves of the first-order model of \cite{liu2010nonlinear}, our second-order analytical quartic solution \eqref{P4}, and the FE computation of quartic indentation in Figure \ref{FIPD-Error}$A$.
Moreover, in Figure \ref{FIPD-Error}$B$, we present the percentage difference between both the model of \cite{liu2010nonlinear} and our second-order analytical quartic solution compared to the FE computation with a quartic indentation. 
We note that Liu's model overestimates the external force compared to the FE simulations, while the second-order analytical parabolic solution slightly underestimates the external force when the indentation depth increases but still closely matches the FE result. 
In particular, according to Figure \ref{FIPD-Error}$B$, the percentage difference of the second-order analytical quartic solution only becomes more than 2\% at the indentation depth $D/R\approx0.8$.
As a comparison, the first-order solution Liu's model exihibits the same level of difference at the indentation depth $D/R\approx0.1$.
In addition, at the maximal indentation depth $D/R=1.0$, Liu's model has more than a 7\% difference, which is twice as big as the difference found than when using our second-order analytical quartic solution.

\subsubsection*{Spherical indentation}

\begin{figure}[h]
\begin{center}
\includegraphics[width=1\textwidth]{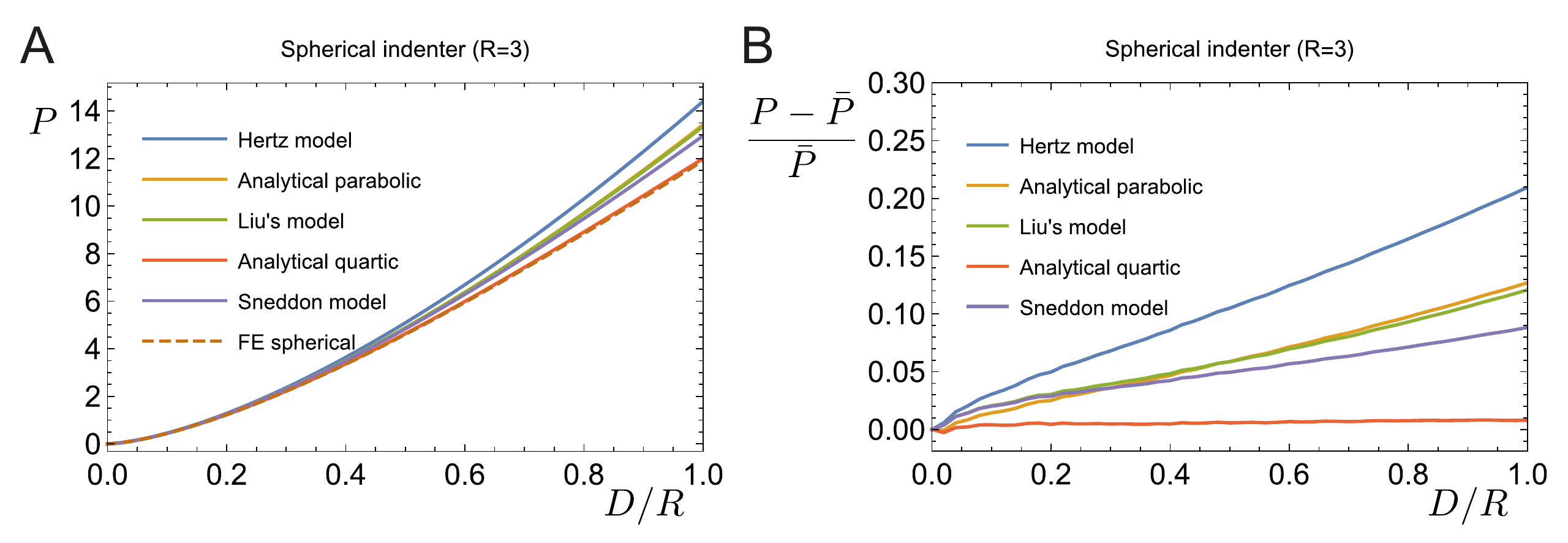}
\caption{Comparisons among the Hertz model, the second-order analytical parabolic solution, Liu's model, the second-order analytical quartic solution, Sneddon model, and the FE simulation result of the spherical indentation. (A) The corresponding force-displacement curves; (B) the percentage difference of the Hertz model, the second-order analytical parabolic solution, Liu's model, the second-order analytical quartic solution, and Sneddon model, to simulation result of the spherical indentation, where $\bar{P}$ is the force by FE simulation.}
\label{SIPD-Error}
\end{center}
\end{figure}

Finally, for the more common indentation problem involving a spherical indenter, we show the force-displacement curves of all indentation models listed in Table \ref{table1} and the FE spherical indentation in the Figure \ref{SIPD-Error}$A$.
In this case, all the indentation models significantly overestimate the applied forces, except for the second-order analytical quartic solution which gives a nearly perfect prediction.
In Figure \ref{SIPD-Error}$B$, we present the percentage differences of all indentation models listed in Table \ref{table1} compared to the FE spherical indentation.
The first-order Hertz model with parabolic indenter exhibits the biggest difference over the whole indentation process. The difference is more than 5\% for the indentation depth $D/R\approx0.2$ and over 20\% at the maximal indentation depth $D/R=1.0$.
Liu's model and the second-order analytical parabolic solution make similar predictions. 
Their differences are more than 5\% when the indentation depth $D/R\gtrapprox0.4$ and over 12\% at the maximal indentation depth $D/R=1.0$.
Due to using the accurate spherical indenter profile function, the first-order Sneddon model still provides a better prediction, but the difference is over 5\% when the indentation depth $D/R\gtrapprox0.5$ and is around 9\% at the maximal indentation depth $D/R=1.0$.
Finally, our second-order analytical quartic solution makes the best prediction, where the difference is less than 1\% over the whole indentation process. 
This improvement is partly due to our consideration of second-order deformations but also due to the retention of higher order $(O(\varepsilon^3))$ terms in the expansion of the shape of the indenter tip.
According to Figures \ref{FEM-Different-Indenters}$A$ and \ref{FIPD-Error}$A$, such close agreement is due to the fact that the second-order analytical quartic solution slightly underestimates the FE quartic simulation, while the FE spherical indentation also slightly underestimates the fourth-order indentation.
 
\subsection{Limitations}
In the previous subsection, we have shown that the second-order quartic solution makes a very close prediction of the finite indentation for an incompressible neo-Hookean solid. However, there is still a limitation of this current second-order indentation model in accounting for other incompressible hyperelastic materials, such as the incompressible Mooney-Rivlin solid.

Consider the energy function of an incompressible Mooney-Rivlin solid 
\begin{equation}
W_{MR}=C_{10}(\bar I_1-3)+C_{01}(\bar I_2-3),
\end{equation}
where $C_{10}$ and $C_{01}$ are material constants; $\bar I_1=\mathrm{tr}\left(\boldsymbol{F}^T\boldsymbol{F}\right), ~\bar I_2=\mathrm{tr}\left((\boldsymbol{F}^T\boldsymbol{F})^{-1}\right)$.
Up to the third-order elasticity, it can be expanded as
\begin{equation}
\begin{aligned}
W_{MR}^*=&-(C_{10}+C_{01})J_{2}+\frac{C_{10}+C_{01}}{2}J_1^2-(C_{10}+2C_{01})J_{1}J_{2}
\\
&+\frac{C_{10}+2C_{01}}{3}J_1^3+(C_{10}+2C_{01})J_{3},
\end{aligned}
\end{equation}
where $a_1=-(C_{10}+C_{01})$. Hence, the two material constants can be represented as 
\begin{equation}
C_{10}=-(1-T)a_1,~C_{01}=-T a_1,
\end{equation}
where $0\leq T\leq1$, and for $T=0$ the energy function reduces to the incompressible neo-Hookean material.

\begin{figure}[h]
\begin{center}
\includegraphics[width=0.6\textwidth]{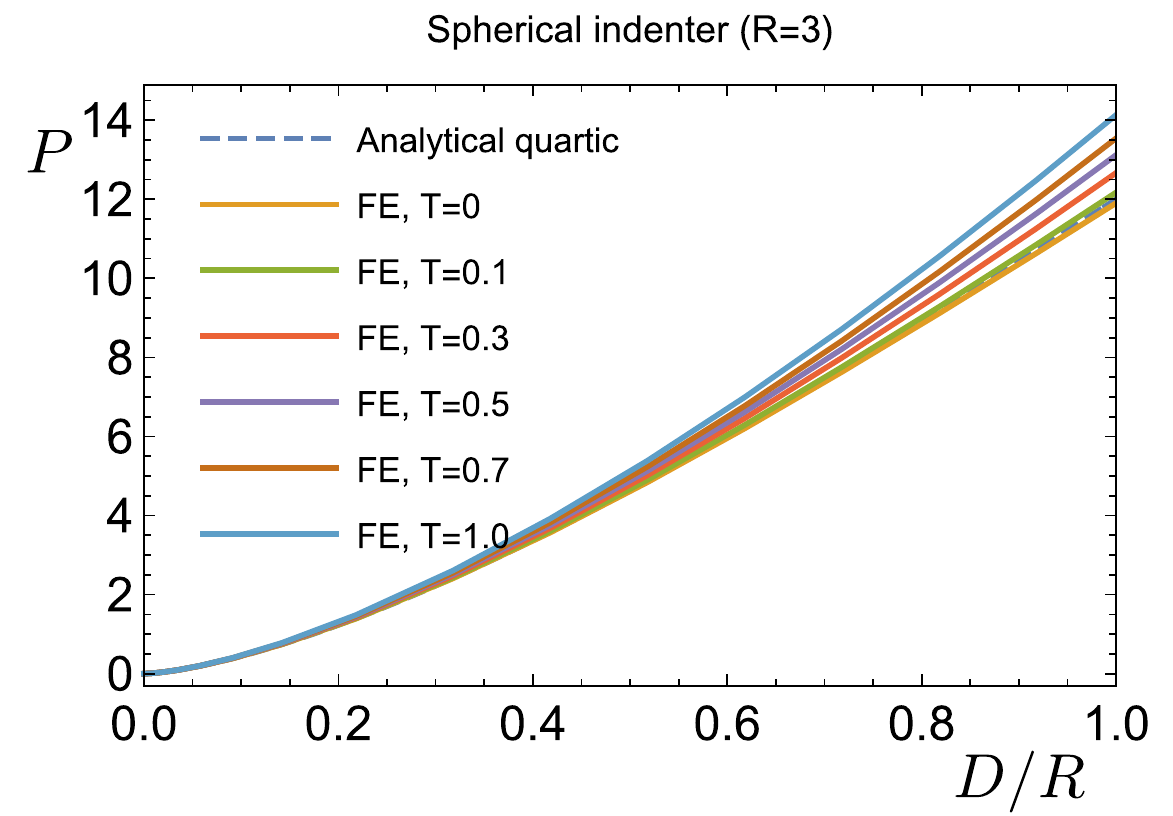}
\caption{Comparison of force-displacement curves among the FE simulation results on Mooney-Rivlin solid (with $T=0,~ 0.1,~ 0.3,~ 0.5,~ 0.7,~ 1.0$) and the second-order analytical quartic solution.}
\label{FEMR}
\end{center}
\end{figure}

In Figure \ref{FEMR}, we present a comparison of the force-displacement curves among the FE simulation results for a Mooney-Rivlin solid (with $T=0$, 0.1, 0.3, 0.5, 0.7, 1.0) and the second-order analytical quartic solution. The force increases as $T$ increases, while the theoretical prediction does not account for the variation of $T$. According to the equations \eqref{D4} and \eqref{P4}, the final expressions of the total displacement and force for an incompressible solid $\eta=0.5$ only involves one material constant $\mu$ or $a_1$. In addition, recall that $a_1=-(C_{10}+C_{01})$.
It is, thus, obvious that this second-order solution would only reflect the sum of $C_{10}$ and $C_{01}$ and can not reflect the separate variation of these two constants.

In other words, the second-order solutions are not available to account for incompressible hyperelastic solids with more than one material constant. Further investigation of this problem shows that the first-order incompressible condition $
\Delta=0$ causes the first-order deformation to go to zero, which further suppresses the material constants $a_3$, $a_4$, $a_5$ in the second-order terms and the final expression of the total force. Hence, we need to consider higher-order elasticity models to overcome this limitation.

\section{Conclusions}

In this paper, we re-present the general solution of the indentation problem using second-order elasticity.
Going beyond the first-order analytical solution provided by Sneddon, the second-order solution can account for the nonlinear deformation and stress during the indentation process. It provides a significant step towards theoretically modelling the indentation problem for hyperelastic materials, especially for the common nano-indentation test on bio-materials.

We have identified and corrected mistakes made by \cite{sabin1983contact} and \cite{giannakopoulos2007spherical}.
We have not only corrected their second-order solutions for parabolic indentation but also have provided the second-order analytical solution for quartic indentation, which is a much better approximation to the spherical indentation (Figure \ref{SIPD-Error}). 
We verify the indentation models by comparing them to FE simulations.
For all indentation models, the second-order solutions show improved predictions compared to the first-order solutions.
In addition, for spherical indentation with an incompressible neo-Hookean material, our second-order quartic solution only exhibits less than 1\% difference for all indentation depths $D$ less than or equal to the indenter radius $R$.
As a comparison, when the indentation depth $D$ equals to the indenter radius $R$, the classic Hertz model exhibits more than 20\% difference, while the corrected second-order parabolic solution exhibits 12\% difference and the first-order Sneddon model nearly 10\% difference.
Consequently, we believe the second-order solutions, especially the second-order quartic solution, should be widely adopted in future experimental and theoretical studies.
The second-order indentation models have limitations in accounting for some high-order incompressible hyperelastic materials. Refining this methodology by introducing the higher-order elasticity and using a better approximation of the spherical indenter profile should further improve prediction.

\section*{Acknowledgement}
This research was funded by EPSRC grants EP/S030875/1 and EP/S020950/1. We also thank Professor Michel Destrade (NUI Galway) for his help with the incompressible limit. Raimondo Penta conducted the research according to the inspiring scientific principles of the national Italian mathematics association Indam (``Istituto nazionale di Alta Matematica”).

\appendix

\renewcommand{\thesection}{A\arabic{section}}   
\renewcommand{\thetable}{A\arabic{table}}   
\renewcommand{\thefigure}{A\arabic{figure}}
\renewcommand{\theequation}{A\arabic{equation}}
\setcounter{figure}{0}
\setcounter{equation}{0}

\section*{Appendix A}\label{sec:counting}
For the parabolic indenter, according to Weber-Sonin-Schafheitlin integral \citep{korenev2002bessel}, the expressions for $F_{i}(r,0)$ and $G_{i}(r,0)$ ($i=0,~1,~2$) are
\begin{equation}
\begin{aligned}
F_0(r,0)=&\begin{cases}
\dfrac{\varepsilon  \left(2 a^2-r^2\right)}{4 a (1-\eta)},  \quad (0\leq r\leq a)\\
\dfrac{\varepsilon }{2 \pi  a (1-\eta)} \left(a \sqrt{r^2-a^2}-\left(r^2-2 a^2\right) \arcsin\left(\dfrac{a}{r}\right)\right), \quad (r > a)\\
\end{cases}\\
F_1(r,0)=&\begin{cases}
\dfrac{2 \varepsilon \sqrt{a^2-r^2}}{\pi  a\left(1- \eta \right)}, \quad (0\leq r\leq a) \\
0,  \quad (r > a) \\
\end{cases}\\
F_2(r,0)=&\begin{cases}
\dfrac{\varepsilon}{a(1-\eta)}, \quad (0\leq r\leq a)\\
\dfrac{2\varepsilon}{\pi  a\left(1- \eta \right) }\left(\arcsin\left(\dfrac{a}{r}\right)-\dfrac{a}{\sqrt{r^2-a^2}}\right), \quad (r > a) \\
\end{cases}\\
G_0(r,0)=&\begin{cases}
\dfrac{2 \varepsilon}{3 \pi r a (1-\eta)} \left(a^3-\left(a^2-r^2\right)^{3/2}\right), \quad (0\leq r\leq a)\\
\dfrac{2  \varepsilon a^3}{3 \pi  r a(1- \eta)}, \quad (r > a) \\
\end{cases}\\
G_1(r,0)=&\begin{cases}
\dfrac{\varepsilon r}{2 a(1-\eta)}, \quad (0\leq r\leq a)\\
\dfrac{\varepsilon}{\pi  r a (1-\eta)}\left(r^2 \arcsin\left(\dfrac{a}{r}\right)-a \sqrt{r^2-a^2}\right), \quad (r > a) \\
\end{cases}\\
G_2(r,0)=&\begin{cases}
\dfrac{2 \varepsilon r}{ \pi a (1- \eta)\sqrt{a^2-r^2}}, \quad (0\leq r\leq a)\\
0, \quad (r > a) \\
\end{cases}
\end{aligned}
\end{equation}

\newpage
\section*{Appendix B}
For the quartic indenter, according to Weber-Sonin-Schafheitlin integral \citep{korenev2002bessel}, the expressions for $F_{i}(r,0)$ and $G_{i}(r,0)$ ($i=0,~1,~2$) are
\begin{equation}
\begin{aligned}
F_0(r,0)=&\begin{cases}
\dfrac{8 a^4 \varepsilon  \left(\varepsilon ^2+3\right)-12 a^2 r^2 \varepsilon -3 r^4 \varepsilon ^3}{48 a^3 (1-\eta)}, \quad(0\leq r\leq a)\\
\dfrac{\varepsilon}{24 \pi  a^3 (1-\eta)}  \left\{a \sqrt{r^2-a^2} \left[2 a^2 \left(\varepsilon ^2+6\right)+3 r^2 \varepsilon ^2\right]\right. \\
\left.+\left[8 a^4 \left(\varepsilon ^2+3\right)-12 a^2 r^2-3 r^4 \varepsilon ^2\right] \arcsin\left(\frac{a}{r}\right)\right\}, \quad (r > a)\\
\end{cases}\\
F_1(r,0)=&\begin{cases}
\dfrac{2 \varepsilon  \sqrt{a^2-r^2} \left[a^2 \left(2 \varepsilon ^2+9\right)+4 r^2 \varepsilon ^2\right]}{9 \pi  a^3 (1-\eta)}, \quad (0\leq r\leq a) \\
0,  \quad (r > a) \\
\end{cases}\\
F_2(r,0)=&\begin{cases}
\dfrac{\varepsilon  \left(a^2+r^2 \varepsilon ^2\right)}{a^3 (1-\eta)}, \quad (0\leq r\leq a)\\
\dfrac{2 \varepsilon}{3 \pi  a^3 (1-\eta ) }  \left\{3 \left(a^2+r^2 \varepsilon ^2\right) \arcsin \left(\frac{a}{r}\right)\right.\\
\left. +a \left(r^2-a^2\right)^{-1/2} \left[a^2 \left(\varepsilon ^2-3\right)-3 r^2 \varepsilon ^2\right]\right\}, \quad (r > a) \\
\end{cases}\\
G_0(r,0)=&\begin{cases}
\dfrac{\varepsilon}{144 \left(1-\eta\right)}  \left\{\dfrac{32 \left(2 \varepsilon ^2+3\right)}{\pi  a r}\left[a^3-\left(a^2-r^2\right)^{3/2}\right]\right.\\
\left.+\dfrac{3 \varepsilon ^2 \left(8 a^2 r^2-8 a^4-3 r^4\right)}{a^3}\right\}, \quad (0\leq r\leq a)\\
\dfrac{\varepsilon}{72 \pi  a^3  \left(1-\eta\right) r}\left\{16 a^5  \left(2 \varepsilon ^2+3\right)
+9a r \varepsilon ^2\sqrt{r^2-a^2}\left(r^2-2a^2\right)\right.\\
\left.-3 r \varepsilon ^2 \left(8 a^4-8 a^2 r^2+3 r^4\right) \arcsin\left(\frac{a}{r}\right)\right\}, \quad (r > a) \\
\end{cases}\\
G_1(r,0)=&\begin{cases}
\dfrac{r \left(2 a^2 \varepsilon +r^2 \varepsilon ^3\right)}{4 a^3 \left(1- \eta \right)}, \quad (0\leq r\leq a)\\
\dfrac{\varepsilon}{6 \pi  a^3 r \left(1- \eta \right)} \left\{ 3 r^2 \left(2 a^2  +r^2 \varepsilon ^2\right) \arcsin\left(\frac{a}{r}\right)-3 a r^2 \varepsilon ^2 \sqrt{r^2-a^2}\right.\\
\left.-2 a^3 \left(\varepsilon ^2+3\right) \sqrt{r^2-a^2} \right\}, \quad (r > a) \\
\end{cases}\\
G_2(r,0)=&\begin{cases}
\dfrac{8 r^3 \varepsilon ^3-a^2 r \left(4 \varepsilon ^3-6 \varepsilon \right)}{3 \pi  a^3 \left(1- \eta \right) \sqrt{a^2-r^2}}, \quad (0\leq r\leq a)\\
0, \quad (r > a) \\
\end{cases}
\end{aligned}
\end{equation}

\newpage
\section*{Appendix C}
\begin{equation}
\begin{aligned}
I_{1}=&\int_{t}^{\infty} \ln \left[\sqrt{\xi^{2}-a^{2}}+\sqrt{\xi^{2}-t^{2}}\right] \frac{\mathrm{d} \xi}{\xi \sqrt{\xi^{2}-a^{2}}}, \quad t \geq a .\\
I_{2}=&\int_{t}^{\infty} \frac{\sqrt{\xi^{2}-t^{2}}}{\xi \sqrt{\xi^{2}-a^{2}}} \arcsin\left(\dfrac{a}{\xi}\right) \mathrm{d} \xi, \quad t \geq a . \\
I_{3}=&\int _{0}^{a}\dfrac{\ln\sqrt{a^{2} -t^{2}}}{\sqrt{r^{2} -t^{2}}} \mathrm{d}t \quad t \leq a , \quad I_{4}=\int_r^{\infty} \dfrac{1}{\xi}\arcsin \left(\dfrac{a}{\xi}\right) \mathrm{d} \xi.\\
I_{5}=&\int_{r}^{a} \ln \left(\frac{a+t}{a-t}\right) \frac{\mathrm{d} t}{\sqrt{t^{2}-r^{2}}} \\
I_{6}=&\frac{1}{2} \int_{r}^{a}\left(1-t^{2} / a^{2}\right) \ln \left(\frac{a+t}{a-t}\right) \frac{\mathrm{d} t}{\sqrt{t^{2}-r^{2}}}\\
I_{7}=& \int_{a}^{\infty}\left[\frac{x}{a} \arcsin(a / x)-\frac{\sqrt{x^{2}-a^{2}}}{x}\right] \\
& \times\left[\arcsin(a / x)-\frac{a}{\sqrt{x^{2}-a^{2}}}\right] \arcsin\left(\sqrt{\frac{a^{2}-r^{2}}{x^{2}-r^{2}}}\right) \mathrm{d} x, \quad r \leqq a.\\
I_8=&\int_a^\infty\left\{\frac{32 a_1 \varepsilon ^2  \arcsin\left(\sqrt{\frac{r^2-a^2}{r^2-\xi ^2}}\right)}{\pi ^3 a^2 (\eta -1) \xi }\left[\frac{a^3-2 a \xi ^2}{\sqrt{\xi ^2-a^2}} \arcsin\left(\frac{a}{\xi }\right)+a^2 +\xi ^2  \arcsin\left(\frac{a}{\xi }\right)^2\right]\right.\\
&\left.+\frac{16 a_1 \varepsilon ^4 \arcsin\left(\sqrt{\frac{r^2-a^2}{r^2-\xi ^2}}\right)}{3 \pi ^3 a^4 (\eta -1) \xi} \left[9 a^2 \xi ^2 +9 \xi ^4\arcsin\left(\frac{a}{\xi }\right)^2+\frac{2 a^5+9 a^3 \xi ^2-18 a \xi ^4}{ \sqrt{\xi ^2-a^2}} \arcsin\left(\frac{a}{\xi }\right)\right]\right.\\
&\left.+\frac{16 a_1 \varepsilon ^6 \arcsin\left(\sqrt{\frac{r^2-a^2}{r^2-\xi ^2}}\right)}{9 \pi ^3 a^6 (\eta -1) \xi} \left[9 \xi ^6  \arcsin\left(\frac{a}{\xi }\right)^2-a^2 \left(2 a^4-3 a^2 \xi ^2-9 \xi ^4\right)\right.\right.\\
&\left.\left.+6 a \xi ^2 \frac{a^4+a^2 \xi ^2-3 \xi ^4}{\sqrt{\xi ^2-a^2}} \arcsin\left(\frac{a}{\xi }\right)\right]\right\}\mathrm{d}\xi.
\end{aligned}
\end{equation}

\newpage
\section*{Appendix D: FEM validation}
\subsubsection*{Influence of the size effect}
As shown in Figure \ref{FE-Models}, we establish FE models for parabolic, quartic, and spherical indentations with $R=3\text{mm}$ and different scales of the substrate $r=h=60 \text{mm},~90 \text{mm},~180\text{mm},~270\text{mm}$, respectively.
At the same time, we set the same boundary conditions $u_{Bz}=0$ for all FE models, where $u_{Bz}$ represents the displacement of the bottom surface along the $z$ direction.

\begin{figure}[h]
\begin{center}
\hspace*{0cm} 
\includegraphics[width=1\textwidth]{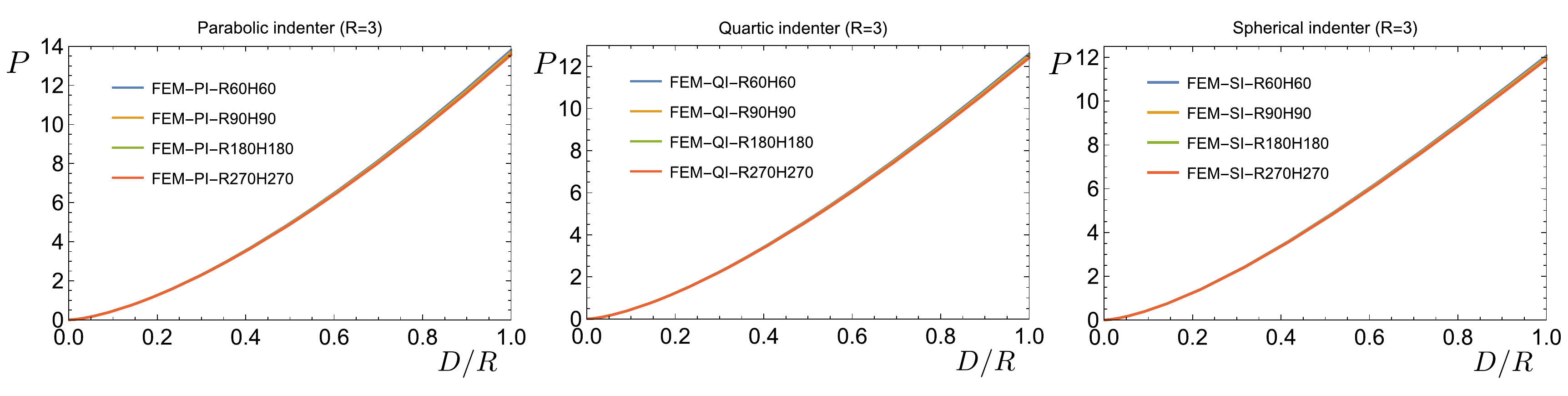}
\caption{The force-displacement curves obtained from the FE simulations of parabolic, quartic, and spherical indentations, with different sizes.}
\label{FEM-Size-Combined}
\end{center}
\end{figure}

Figure \ref{FEM-Size-Combined} shows the force-displacement curves obtained from the FE simulations of parabolic, quartic, and spherical indentation, respectively. For all subfigures, there are small but visible differences between curves for the minimal size $r=h=60\text{mm}$ and maximal size $r=h=270\text{mm}$.
As the size increases, the curves for the size $r=h=180\text{mm}$ and $r=h=270\text{mm}$ nearly overlap, which indicates that the simulation for this case would approach a convergent result as the cylinder size $r=h\geq180\text{mm}$. 
Further increasing the size of the cylinder will not create a significant difference compared to the real half-space.
Hence, for this indentation problem, a cylinder with the size $r=h=270\text{mm}$ should be big enough to mimic the half-space body.

\subsubsection*{Influence of the boundary conditions}

Then, we establish FE models for parabolic, quartic, and spherical indentations with indenter radius $R=3\text{mm}$ and cylinder size $r=h=270\text{mm}$ obtained previously.
To study the influence of the boundary conditions, we set different boundary conditions at the bottom surface and lateral surface of the cylindrical substrate, which are $u_{Bz}=0$, $u_{Bz}=u_{Lr}=0$, $u_{Bz}=u_{Lz}=0$, $u_{Bz}=u_{Lr}=u_{Lz}=0$, $u_{Br}=u_{Bz}=0$, $u_{Br}=u_{Bz}=u_{Lr}=0$, $u_{Br}=u_{Bz}=u_{Lz}=0$, and $u_{Br}=u_{Bz}=u_{Lr}=u_{Lz}=0$, where the subscript $u_{Br}$ and $u_{Bz}$ indicate the displacement of the bottom surface along the $r$ and $z$ directions, $u_{Lr}$ and $u_{Lz}$ indicate the displacement of the lateral surface along the $r$ and $z$ directions.

\begin{figure}[h]
\begin{center}
\hspace*{0cm} 
\includegraphics[width=1\textwidth]{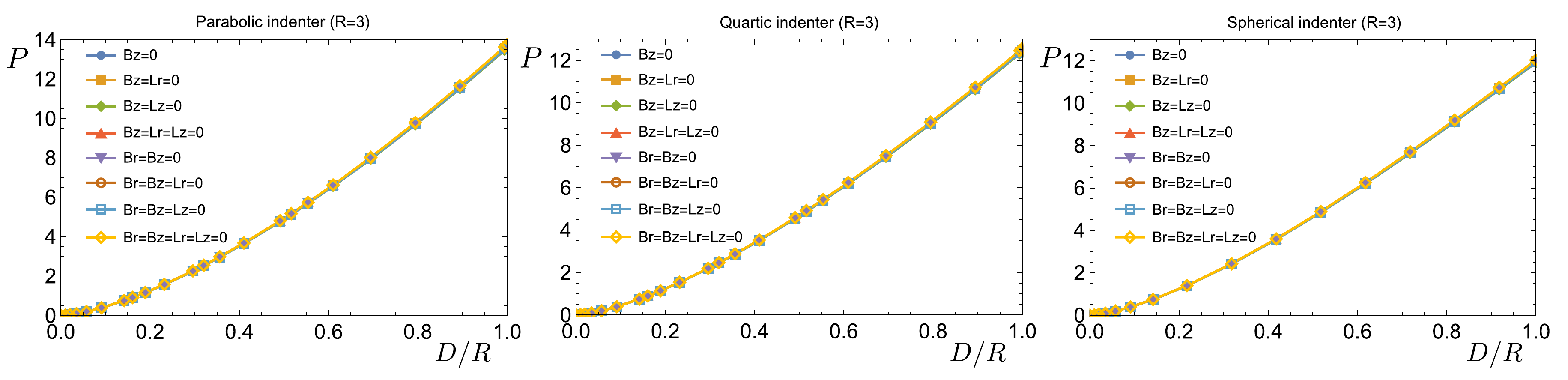}
\caption{The force-displacement curves obtained from the FE simulations of parabolic, quartic, and spherical indentations, with different boundary conditions.}
\label{FEM-BCs-Combined}
\end{center}
\end{figure}

Figure \ref{FEM-BCs-Combined} shows the force-displacement curves obtained from FE simulations of parabolic, quartic, and spherical indentation, respectively. 
As shown in all subfigures, the force-displacement curves under different boundary conditions are not significantly different.
In other words, as there are no boundary conditions to be specified for the real half-space body when the cylinder size is big enough, the influence of the boundary conditions on this indentation problem is negligible.

Hence, based on the studies of these two influential factors above, we could then trust the FE simulations that use a cylinder with the size $r=h=270\text{mm}$ and the boundary condition $u_{Bz}=0$ to represent the half-space substrate.

\bibliographystyle{unsrtnat}
\bibliography{indentation}  






\end{document}